\documentclass[preprint]{elsarticle}
\usepackage[utf8]{inputenc}
\usepackage{amsmath}
\usepackage{amsfonts}
\usepackage{amssymb}
\usepackage{graphicx}
\usepackage{subcaption}
\usepackage{fullpage}
\usepackage{float}
\usepackage{xcolor}
\usepackage{enumitem}
\usepackage[colorlinks,allcolors=blue]{hyperref}

\DeclareMathOperator\erf{erf}
\DeclareMathOperator\erfc{erfc}
\journal{Journal of Computational Physics}

\begin{document}

\begin{frontmatter}

\title{A semi-analytical transient undisturbed velocity correction scheme for wall-bounded two-way coupled Euler-Lagrange simulations}

\author[affil1]{Akshay Chandran}
\author[affil2]{Fabien Evrard}
\author[affil1]{Berend van Wachem\corref{cor1}}
\ead{berend.vanwachem@ovgu.de}

\address[affil1]{Chair of Mechanical Process Engineering, Otto-von-Guericke-Universit\"at Magdeburg, Universit\"atsplatz 2, 39106 Magdeburg, Germany}
\address[affil2]{Department of Aerospace Engineering, University of Illinois Urbana-Champaign, 104 South Wright Street, Urbana, IL 61801, United States of America}

\cortext[cor1]{Corresponding author: }

\date{}

\begin{abstract}
Force closure models employed in Euler-Lagrange (EL) point-particle simulations rely on the accurate estimation of the 
undisturbed fluid velocity at the particle center to evaluate the fluid forces on each particle. Due to the 
self-induced velocity disturbance of the particle in the fluid, two-way coupled EL simulations only have access to the disturbed velocity. 
The undisturbed velocity can be recovered if the particle-generated disturbance is estimated.
In the present paper, we model the velocity disturbance generated by a regularized forcing near a planar wall, which, 
along with the temporal nature of the forcing, provides an estimate of the unsteady velocity disturbance of 
the particle near a planar wall. We use the analytical solution for a singular in-time transient Stokeslet near a planar wall 
(\citet{Felderhof2009}) and derive the corresponding time-persistent Stokeslets. The velocity disturbance due to a regularized 
forcing is then obtained numerically via a discrete convolution with the regularization kernel. The resulting Green's functions for 
parallel and perpendicular regularized forcing to the wall are stored as pre-computed temporal correction maps. By storing the 
time-dependent particle force on the fluid as fictitious particles, we estimate the unsteady velocity disturbance 
generated by the particle as a scalar product between the stored forces and the pre-computed Green's functions. 
Since the model depends on the analytical Green's function solution of the singular Stokeslet 
near a planar wall, the obtained velocity disturbance exactly satisfies the no-slip condition and does not 
require any fitted parameters to account for the rapid decay of the disturbance near the wall. The numerical evaluation of 
the convolution integral makes the present method suitable for arbitrary regularization kernels. Additionally, the 
generation of parallel and perpendicular correction maps enables to estimate the velocity disturbance due to particle motion 
in arbitrary directions relative to the local flow. The convergence of the method is studied on a fixed particle 
near a planar wall, and verification tests are performed on a settling particle parallel to a wall and a free-falling 
particle perpendicular to the wall.
\end{abstract}

\end{frontmatter}

\section{Introduction}
    
    Particle-laden flows are governed by the intricate interplay between fluid and particulate matter, and are ubiquitous in both natural 
    and industrial settings. From natural phenomena such as the motion of droplets in clouds and sediment transport in rivers \cite{Zhang2015}, 
    to industrial applications like powder processing in manufacturing, understanding the dynamics of these multiphase flows 
    is of paramount importance. The wide range of length and time scales present in such flows necessitates 
    a variety of numerical methods for their scientific study. One class of numerical methods uses a continuum description 
    of the fluid while describing particle dynamics using Newton’s second law; these methods are termed Euler-Lagrange (EL) methods.
    EL point-particle methods do not fully resolve the fluid flow around the particle and therefore do not impose boundary conditions 
    on the particle surfaces. Such methods rely on empirical closures to model the momentum exchange 
    between the two phases. When the particle mass loading and/or volume fractions are significant, the momentum transferred from the  
    particle phase into the fluid cannot be neglected, and is accounted for through the inclusion of a source term in the fluid momentum equations. 
    Methods where the particle presence in the fluid is communicated through a source term contribution in the fluid momentum equations 
    are known as ``two-way" coupled EL methods \cite{Balachandar2010,Wachem2022}.

    The fundamental challenge in using force closure models for point-particle simulations is to obtain the \textit{undisturbed fluid velocity} 
    at the center of the particle. The undisturbed fluid velocity is defined as what the local fluid velocity would have been if the particle 
    would not have been present.
    For example, for a particle settling in a quiescent flow, the undisturbed velocity is zero. In two-way coupled 
    EL simulations, the presence of the particle disturbs the local flow around it, preventing direct access to the undisturbed velocity. 
    Instead, only the \textit{disturbed fluid velocity} obtained is easily accessible. Using the 
    disturbed fluid velocity in the force closure models leads to underestimation of the particle-fluid interaction force, thereby leading to inaccurate prediction of the particle dynamics. Recovering the 
    undisturbed velocity from the disturbed fluid velocity is typically done by subtracting the disturbance flow generated by the particle. 
    The velocity disturbance in the fluid 
    has different spatial and temporal characteristics depending on nondimensional quantities such as the Stokes number 
    ($\mathrm{St} = \tau_{\mathrm{p}}/\tau_{\mathrm{f}}$) and particle Reynolds number. Here, $\tau_{\mathrm{p}}$ and $\tau_{\mathrm{f}}$ are 
    the particle and fluid response time, respectively, to an imposed acceleration. When the particle Reynolds number is large, 
    the particle reaction force on the fluid is relatively small, 
    resulting in smaller magnitudes of the velocity disturbance and which is convected with the background flow. 
    Similarly, when the particle responds slower to 
    an imposed acceleration than the fluid (i.e., $\tau_{\mathrm{p}} > \tau_{\mathrm{f}}$), the velocity disturbance generated by the particle in the 
    fluid diffuses long enough within the particle time scale, so that the particle experiences a nearly steady state velocity disturbance. In contrast, 
    when $\tau_{\mathrm{p}} < \tau_{\mathrm{f}}$, the velocity disturbance remains transient within the particle time scale. In the former case, 
    a steady disturbance model suffices \cite{Balachandar2019,Evrard2020a}, whereas in the latter, a transient disturbance model is necessary to accurately 
    predict the disturbance velocity as a function of time.
    In the presence of a wall, this disturbance velocity must also satisfy the no-slip condition on the wall. Analytical and 
    empirical correlations are typically used to model this steady or transient disturbance flow \cite{Gualtieri2015,Balachandar2019,Pakseresht2020a,Pakseresht2021,Horwitz2022,Balachandar2022}.

	Numerous analytical and empirical frameworks have been proposed to estimate the undisturbed velocity. 
	\citet{Gualtieri2015} introduced a correction scheme based on a closed-form solution of 
	the vorticity disturbance equation, which diffuses over a short time. \citet{Horwitz2016} 
	employed a power series expansion with empirical coefficients to model the spatial variation of the 
	velocity disturbance. Their model requires calibration depending on the type of solver used and is not 
	generalized for arbitrary particle diameter to mesh-spacing ratios. An extension of their model to finite Reynolds 
	numbers was subsequently proposed \citep{Horwitz2018}. \citet{Ireland2017} model the 
	velocity disturbance based on the solution of the steady Stokes equations. However, their solution is restricted 
	to low Reynolds numbers, and does not account for the temporal nature of the velocity disturbance. 
	\citet{Evrard2020a} model the 
	velocity disturbance for an arbitrary particle diameter to mesh-spacing ratio using the Stokes flow 
	solution through a regularized momentum source, with extensions to finite Reynolds numbers based 
	on the Oseen flow equations. However, their solution does not account for the transient nature of the 
	disturbance, and is limited to high Stokes numbers.
	\citet{Balachandar2019} developed an empirical model that accounts for the unsteady nature of 
	the velocity disturbance across a range of particle Reynolds numbers. Their model accomodates flows 
	in both the Stokes regime and at finite Reynolds numbers using different weighted kernels that 
	describe the temporal history of the force variation. However, their model does not consider the 
	presence of a wall.
	
	Few solutions have been proposed to obtain the velocity disturbance near a wall.
    \citet{Pakseresht2020a} obtained the disturbance generated by an object with dimensions equivalent to a computational cell near a 
    planar wall and evaluated the velocity disturbance at its location. They model the velocity disturbance generated by an infinitesimal force at the 
    cell center subjected to Stokes drag. Empirical factors are employed to account for the presence of the wall, shape of the cell, inertial effects, 
    and the transient nature of the disturbance. 
    \citet{Pakseresht2021} numerically solve an additional velocity disturbance governing equation 
    that approximates the pressure gradient contribution by its value in the Stokes regime. Their model requires solving an additional 
    discrete equation on a smaller computational grid and is ideally suited for 
    the dilute regime. Extension to the dense regime necessitates the solution of the discrete velocity disturbance equations 
    around every particle.
    \citet{Horwitz2022} numerically obtained the discrete Green's function (DGF) of the steady Stokes operator 
    in a channel geometry. Although their model shows promise in highlighting the importance of obtaining the undisturbed fluid velocity in 
    a turbulent channel flow, the generated DGFs rely on the numerical fluid flow solver used to obtain them, and producing 
    unsteady DGFs requires colossal data storage capabilities. \citet{Balachandar2022} used a mirror particle across a planar wall to 
    obtain the volume-corrected velocity disturbance of a particle. But such an approach does not satisfy a no-slip condition or a uniform 
    slip velocity over the entire wall. Furthermore, the decay of the velocity near the wall is not physical.
    For the simplest case of a steady Stokeslet, apart from a single reflection of the Stokeslet, higher-order solutions of the Stokes 
    equations become necessary to satisfy the no-slip condition
    \cite{Blake1971,Cortez2015}. Additionally, \citet{Balachandar2022} use an empirical model to account for the 
    unsteady nature of the velocity disturbance.
    
    In the current paper, we introduce a new model for correcting the self-induced fluid velocity disturbance of a particle that is based on an 
    analytical model of the velocity disturbance generated due to an impulse near a planar wall \cite{Felderhof2009}. The present model is an 
    extension of the transient free-space velocity correction model in \citet{Evrard2024a}. Since the model 
    is derived from the volume-filtered Euler-Lagrange equations, the self-induced velocity disturbance accounts for the local 
    volume-fraction and is, therefore, suitable for dense particle-laden flows. Unlike other methods that rely on empirical 
    models to determine the velocity disturbance near the wall, we present a fully deterministic model that uses no 
    ad-hoc parameters. Additionally, the model is designed to handle arbitrary orientations of the relative velocity 
    and the feedback force.
    
	The remainder of this paper is divided into four parts. Section \ref{sec:mathematical_formulation} discusses the 
	governing equations for the unsteady velocity disturbance near a planar wall and describes how we obtain the corresponding 
	velocity disturbance 
	due to a persistent regularized forcing. In Section \ref{sec:numerical_methodology}, we present a methodology for 
	numerically obtaining the velocity disturbance due to a regularized forcing. Model convergence and verifications 
	are then tested in Section \ref{sec:results} for a fixed particle in a laminar boundary layer, a settling particle 
	parallel to the wall, and a free-falling particle normal to the wall. Finally, in Section \ref{sec:summary}, we 
	provide our conclusions and summarize the work. We provide a detailed derivation of the persistent transient 
	Stokeslets and their asymptotic limits in Section \ref{subsec:appendix_persistentstokeslets_verticalforcing} 
	and \ref{subsec:appendix_persistentstokeslets_horizontalforcing}, and the central values of these Stokeslets 
	obtained using an analytic convolution with a Top-hat filter in Section \ref{subsec:centralvalue_regstokeslets}.

\section{Mathematical formulation}
\label{sec:mathematical_formulation}

\quad In Euler-Lagrange (EL) simulations, the motion of a rigid spherical particle in a fluid can be described by Newton's second law,
\begin{equation}
\rho_{\mathrm p} V_{\mathrm{p}} \frac{\mathrm{d} \mathbf{u}_{\mathrm{p}}}{\mathrm{d} t} = \mathbf{F}_{\mathrm{fluid}} + \mathbf{F}_{\mathrm{body}}
\label{eqn:maxey_riley_eqn}
\end{equation}

where $\rho_{\mathrm p}$ is the particle density, $V_{\mathrm p}$ is the volume of the particle, $\mathbf{u}_{\mathrm{p}}$ is the velocity of the center of mass of the particle, $\mathbf{F}_{\mathrm{body}}$ represents the volumetric forces acting on the particle, and $\mathbf{F}_{\mathrm{fluid}}$ is modeled using the Maxey-Riley-Gatignol equations \cite{Maxey1983,Gatignol1983} as the sum of drag, history, added mass, and the undisturbed flow forces acting on the particle due to the relative motion between the particle and the surrounding fluid. Since the surface of the particle is not resolved in a point-particle model, these fluid forces are represented as a function of the 
fluid and particle quantities at the particle center $\mathbf{x}_{\mathrm{p}}$. For example, the fluid drag on a rigid spherical particle of diameter $d_{\mathrm{p}}$ suspended in a fluid of density $\rho_\mathrm{f}$ is given by:
\begin{equation}
\mathbf{F}_{\mathrm{p},\mathrm{drag}} = \frac{3}{4} \frac{\rho_{\mathrm{f}} V_{\mathrm{p}} \mathrm{C}_{\mathrm{D}}}{d_{\mathrm{p}}} \lvert \lvert \tilde{\mathbf{u}}(\mathbf{x}_{\mathrm{p}}) -  \mathbf{u}_{\mathrm{p}} \rvert \rvert ( \tilde{\mathbf{u}}(\mathbf{x}_{\mathrm{p}}) -  \mathbf{u}_{\mathrm{p}} ).
\label{eqn:drag_pointparticle}
\end{equation}

Here $\tilde{\mathbf{u}}(\mathbf{x}_{\mathrm{p}})$ is the \textit{undisturbed fluid velocity} at the particle center, and $\mathrm{C}_{\mathrm{D}}$ is the drag coefficient from experiments 
or numerical simulations \cite{Schiller1933,Zeng2009}. 
 The undisturbed velocity at the center $\mathbf{x}_{\mathrm{p}}$ of the particle denotes the fictitious fluid velocity as if the particle exerted no influence on the background flow. Obtaining this fluid velocity is crucial, since the force closures employed in point-particle models rely on the slip velocity of the particle relative to the fluid flow in the absence of the particle.

In two-way coupled EL simulations, the momentum fed back from the particle disturbs the local flow around it, thereby preventing direct access to $\tilde{\mathbf{u}}(\mathbf{x}_{\mathrm{p}})$. Instead, the fluid velocity interpolated to the particle center is the \textit{disturbed fluid velocity}, $\overline{\mathbf{u}}(\mathbf{x}_{\mathrm{p}})$. $\tilde{\mathbf{u}}$ is commonly evaluated by modeling the self-induced velocity disturbance, $\mathbf{u}^{\prime}$, of the particle and subtracting this from $\overline{\mathbf{u}}$.

\begin{equation}
\tilde{\mathbf{u}} = \overline{\mathbf{u}} - \mathbf{u}^{\prime}.
\label{eqn:disturbed_undist_velocity_relationship}
\end{equation}

The accuracy of the undisturbed velocity estimation and therefore the fluid forces on the particle, relies heavily on the mathematical model 
used to describe the self-induced velocity disturbance of the particle. Additionally, for a particle located near a wall, this disturbance velocity needs to satisfy the 
appropriate wall boundary conditions.


\subsection{Velocity disturbance governing equations}
\label{subsec:velocity_disturbance_governing_equations}

\begin{figure}[ht]
\centering
\includegraphics[scale=1.0]{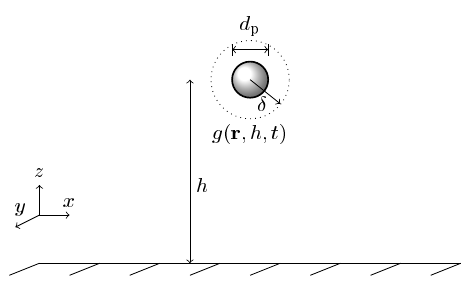}
\caption{Schematic of a particle with diameter $d_{\mathrm{p}}$ located near a planar wall at a distance $h$. Here, $g(\mathbf{r},h,t)$ represents the regularization kernel of the applied forcing with a support radius of $\delta$ denoted by the dotted circle.}
\label{fig:fixedparticle_nearplanarwall}
\end{figure}

\quad To model the velocity disturbance, we assume $\mathbf{u}^{\prime}$ varies on much smaller time and length scales than $\tilde{\mathbf{u}}$, and 
the Reynolds number based on $\mathbf{u}^{\prime}$ is small, i.e., $\rho_{\mathrm{f}} \lvert \mathbf{u}^{\prime} \rvert \delta/\mu_{\mathrm{f}} \ll 1$, for a fluid of dynamic viscosity $\mu_{\mathrm{f}}$, and a force regularization length scale of $\delta$. Consequently, spatial gradients of $\tilde{\mathbf{u}}$ are not advected with $\mathbf{u}^{\prime}$, and vice-versa.
This assumption, in a reference frame moving with the particle, leads to a linear set of governing equations describing the dynamics of $\mathbf{u}^{\prime}$ in the form of the unsteady Stokes equations \cite{Evrard2024a}.
The velocity disturbance generated in the fluid by a momentum source, $\mathbf{F}$, at a distance $h$ away from the wall is then obtained as the solution of the unsteady Stokes equations for an incompressible fluid.

\begin{equation}
\rho_\mathrm{f} \frac{\partial \mathbf{u}^{\prime}}{\partial t} = - \nabla p + \mu_\mathrm{f} \nabla^2 \mathbf{u}^{\prime} + \mathbf{F} g(\mathbf{r},h,t), \quad \nabla \cdot \mathbf{u}^{\prime} = 0.
\label{eqn:unsteady_stokeseqn}
\end{equation}

Here, $\mathbf{r}$ and $t$ are independent variables denoting the space and time coordinates respectively. $p$ represents the pressure, and $g(\mathbf{r},h,t)$ is the normalized force regularization kernel. Figure \ref{fig:fixedparticle_nearplanarwall} shows a sketch of a particle of diameter $d_{\mathrm{p}}$ located at a distance $h$ away from the wall, and whose force on the fluid is regularized using a kernel $g(\mathbf{r},h,t)$ with a compact support of radius $\delta$. For a detailed derivation of the 
linear disturbance governing equations in the volume-filtered Euler-Lagrange framework, the reader is referred to \citet{Evrard2024a}.

For a singular forcing regularization in space and time, i.e., $g(\mathbf{r},h,t) = \zeta(\mathbf{r}) ~\zeta(t)$ and when $h \to \infty$, the solution of Equation \eqref{eqn:unsteady_stokeseqn} is the free-space transient singular Stokeslet \cite{Pozrikidis1992}. Here, $\zeta$ represents the Dirac-delta function. This velocity disturbance, derived in an unbounded domain generated by a singular forcing decays to zero only as $\lvert \mathbf{r} \rvert \to \infty$. For a finite $h$, the velocity disturbance has to decay rapidly to zero as $z \to 0$. Hence, using the free-space transient stokeslet to model $\mathbf{u}^{\prime}$ in the presence of no-slip walls would result in a spurious slip-velocity at the wall. When the equations governing $\mathbf{u}^{\prime}$ are the steady Stokes equations, the method of images can be used to satisfy the no-slip and no-penetration condition at the wall \cite{Blake1971,Cortez2015}.

In the presence of a planar no-slip wall, \citet{Felderhof2009} provides an analytical solution in the half-space $z > 0$ for Equation \eqref{eqn:unsteady_stokeseqn}, when $g(\mathbf{r},h,t) = \delta(\mathbf{r}) ~\delta(t)$ for finite $h$. The solution is obtained via the method of Green's functions for the Fourier transformed unsteady Stokes operator. The Green's functions are obtained in the 2D Fourier space for the wavenumbers $k_x, k_y$ corresponding to the real space coordinates $x, y$ parallel to the planar wall. The corresponding real space solution is obtained numerically using a Hankel transform as follows \cite{Arfken2013,Felderhof2009},

\begin{equation}
G_{ij} (R,z,h,t) = \int_{0}^{\infty} \hat{G}_{ij} (k,z,h,t) J_{n}(kR) ~k ~\mathrm{d}k,
\label{eqn:hankel_transform}
\end{equation}

where $\hat{G}_{ij}$ is the Green's function in fourier space, $k = (k_x^2 + k_y^2)^{1/2}$ is the radial distance in fourier space, $R = (x^2 + y^2)^{1/2}$ is the real space radial distance parallel to the wall from the center of the particle, and $J_{n}(x)$ is the Bessel function of the first kind of order $n$ (see Sections \ref{subsec:appendix_persistentstokeslets_verticalforcing} and \ref{subsec:appendix_persistentstokeslets_horizontalforcing} in the Appendix). The evaluation of this integral can be expensive, as the Bessel functions constitute a slowly converging series sum, and using a truncated series expansion can be computationally demanding. In the present study, this calculation was expedited by employing a fast Bessel function algorithm \cite{Tumakov2019}.

\begin{figure}[t]
\includegraphics[scale=0.9]{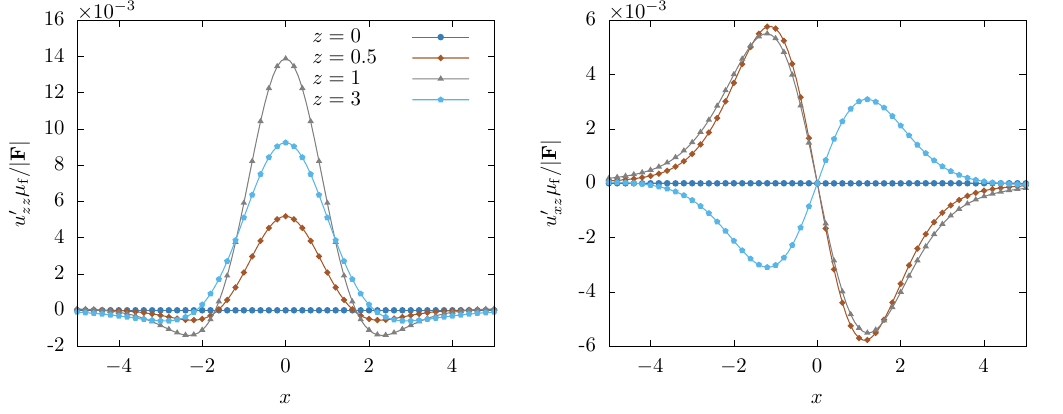}
\caption{Horizontal component ($u^{\prime}_{xz}$), and 
vertical component ($u^{\prime}_{zz}$) of the velocity disturbance from \citet{Felderhof2009} in the $xz$ plane 
at various vertical distance $z$ from the wall for a singular forcing $\mathbf{F} = 4\pi \mathbf{e}_z$ at the point 
$\mathbf{r}_0 = (0,0,1)$ after a time $t=0.5$ since the application of the force. Here the first index in the subscript denotes 
the direction of the velocity, and the second index denotes the direction of forcing.}
\label{fig:singular_velocity_felderhof}
\end{figure}

For a particle located at a finite distance from the wall, the solution of Equation \eqref{eqn:unsteady_stokeseqn} can be found for two distinct cases. When the forcing is parallel to the wall, i.e., $\mathbf{F} = \lVert \mathbf{F} \rVert \mathbf{e}_x$, there are three components of the Green's functions $G_{xx}$, $G_{yx}$, and $G_{zx}$, corresponding to the $u_x^{\prime}$, $u_y^{\prime}$, and $u_z^{\prime}$ velocity responses, respectively. When the forcing is perpendicular to the wall, i.e. $\mathbf{F} = \lVert \mathbf{F} \rVert \mathbf{e}_z$, the resulting velocity disturbance is axisymmetric, and the Green's function has two components, $G_{zz}$ and $G_{Rz}$, corresponding to the vertical $u_z^{\prime}$ and radial $u_R^{\prime} = ((u_x^{\prime})^2 + (u_y^{\prime})^2)^{1/2}$ components of the velocity disturbance, respectively. Figure \ref{fig:singular_velocity_felderhof} shows the velocity field from \citet{Felderhof2009} in the $xz$ plane generated by a singular forcing $\mathbf{F} = 4\pi \mathbf{e}_z$ at the point $\mathbf{r}_0 = (0,0,1)$ after a time $t = 0.5$ since the application of the force. The velocities $u^{\prime}_{xz}$ and $u^{\prime}_{zz}$ decrease rapidly with decreasing $z$, and at $z = 0$, they satisfy the no-slip and no-penetration condition as indicated by the circle shaped markers. For $G_{ij}$ and $u^{\prime}_{ij}$,
the first subscript index denotes the direction of the velocity, and the second index denotes the direction of the applied forcing.

Particles located near a planar wall experience a lift force in addition to an increased drag force \cite{Segre1962,Segre1962a,Vasseur1976,Cox1977}. More importantly, the drag on the particle can be oriented in arbitrary directions when the particle motion is non-aligned with the flow direction. For arbitrary orientations of the force $\mathbf{F}$, the linearity of the governing equations enables us to superpose the velocity disturbance generated by parallel and perpendicular forcings. For example, for a force of the form $\mathbf{F} = F_x \mathbf{e}_x + F_z \mathbf{e}_z$, the resulting velocities are given by $\mathbf{u}^{\prime} = (u^{\prime}_{xx} + u^{\prime}_{Rz} \cos\phi) \mathbf{e}_x + (u^{\prime}_{yx} + u^{\prime}_{Rz} \sin\phi) \mathbf{e}_y + (u^{\prime}_{zx} + u^{\prime}_{zz}) \mathbf{e}_z$. Here $\mathbf{e}_i$ is the unit vector denoting the Cartesian coordinates, $\phi$ represents the orientation of the vector connecting the evaluation point and the particle center relative to $\mathbf{e}_x$ in the plane parallel to the wall containing the particle center.

The momentum transferred by finite-sized particles into the fluid is distributed in space. $g(\mathbf{r},h,t)$ then has a finite support and a singular representation in space using the Dirac delta function is not suitable. The solution of Equation \eqref{eqn:unsteady_stokeseqn} for a regularized 
forcing is analytically hard to obtain. In the next subsection, we elaborate how we approximate this solution numerically.

\subsection{Regularized persistent forcing near a planar wall}
\label{subsec:regularized_persistent_stokeslet_subsection}

\quad We define a regularized forcing of the form $\mathbf{P}(\mathbf{r},t) = \mathbf{F}(t) g(|\mathbf{r} - \mathbf{r}_0|)$, where $g$ is an 
isotropic regularization kernel such as Gaussian, Wendland, Top-hat, etc. Here $\mathbf{r}_0 = \{r_x, r_y, h \}$ represents the location of the forcing. The kernel of the wall-bounded transient singular in space and time stokeslet is denoted by $\mathbf{G}(\mathbf{r},\mathbf{r}_0,t)$ \cite{Felderhof2009}. 
The velocity response $\mathbf{u}^{\prime}_{0}$ to a singular forcing $\mathbf{P}_{0}(\mathbf{r},t) = \mathbf{F} \delta(\mathbf{r} - \mathbf{r}_0) \delta(t - t_0)$ applied at time $t = t_0$ is then:

\begin{equation}
\mathbf{u}^{\prime}_{0}(\mathbf{r},t) = \mathbf{G}(\mathbf{r},\mathbf{r}_0,t) \cdot \mathbf{P}_{0}(\mathbf{r},t).
\label{eqn:velocity_disturbance_singular_forcing}
\end{equation}

The linearity of Equation \eqref{eqn:unsteady_stokeseqn} permits the representation of the response of a 
regularized forcing as a convolution product of $\mathbf{G}(\mathbf{r},\mathbf{r}_0,t)$ with the 
regularization kernel $g(|\mathbf{r} - \mathbf{r}_0|)$. Mathematically, the velocity field generated 
by a regularized forcing is therefore:

\begin{equation}
\mathbf{u}^{\prime}(\mathbf{r},t) = \mathbf{F} \int_{\mathbb{R}^2 \cup \mathbb{R}^{z+}} g(|\mathbf{x} - \mathbf{r}_0|) \mathbf{G}(\mathbf{r} - \mathbf{x},\mathbf{r}_0,t) ~\mathrm{d}x ~\mathrm{d}y ~\mathrm{d}z,
\label{eqn:velocity_disturbance_regularized_forcing}
\end{equation}

where $\mathbb{R}^2 \cup \mathbb{R}^{z+}$ is the three dimensional space in the half-space $z > 0$, representing the union of the two-dimensional space parallel 
to the wall, $\mathbb{R}^2$, and the one-dimensional space above the wall, $\mathbb{R}^{z+}$. Here, $\mathbf{x}$ represents the coordinates in the $\mathbb{R}^2$ space. Numerical simulation of wall-bounded flows use a non-uniform discretization of the computational domain in the wall-normal direction in order to resolve the flow scales near the wall. 
On a uniform sized mesh, the convolution product in Equation \eqref{eqn:velocity_disturbance_regularized_forcing} can be performed in Fourier space utilizing the Fourier convolution theorem. Additionally, the convolution theorem requires the functions to be translationally 
non-invariant. To make use of this, we split this integral into wall-parallel and wall-normal directions respectively as follows:

\begin{equation}
\mathbf{u}^{\prime}(\mathbf{r},t) = \mathbf{F} \int_{\mathbb{R}^{z+}} \left(\int_{\mathbb{R}^2} g(|\mathbf{x} - \mathbf{r}_0|) \mathbf{G}(\mathbf{r} - \mathbf{x},\mathbf{r}_0,t) ~\mathrm{d}x ~\mathrm{d}y \right) \mathrm{d}z.
\label{eqn:convolution_integral_separated_parallel_perpendicular}
\end{equation}

This form of the integral allows us to evaluate the convolution in $\mathbb{R}^2$ in Fourier space, and a real space convolution is performed along $\mathbb{R}^{z+}$. Using the 
convolution theorem, the inner integral can be simplified to:

\begin{equation}
\mathbf{u}^{\prime}(\mathbf{r},t) = \mathbf{F} \int_{\mathbb{R}^{z+}} \mathcal{F}^{-1} \biggl\lbrace \hat{g} (\mathbf{k},z) \star \hat{\mathbf{G}} (\mathbf{k},h,z,t) \biggr\rbrace ~\mathrm{d}z.
\label{eqn:convolution_integral_separated_fourier}
\end{equation}

Here, $\hat{\mathbf{G}}$ represents the solution provided in \citet{Felderhof2009}, and $\mathcal{F}^{-1}$ represents the 
inverse Fourier transform operator. 
The integrand in Equation \eqref{eqn:convolution_integral_separated_fourier} can be evaluated numerically using 
a discrete Fourier transform (DFT). For a computational grid of dimension $\mathrm{N}$, this evaluation in Fourier space 
using a fast Fourier transform has a complexity of $\mathcal{O}(\mathrm{N} \log \mathrm{N})$, in contrast to  a complexity of 
$\mathcal{O}(\mathrm{N}^2)$ for a real space computation. Additionally, a direct integration along the wall-normal direction 
is performed for the integral in Equation \eqref{eqn:convolution_integral_separated_fourier}. This integration proves 
computationally economical, as it exclusively spans the extents of the regularization diameter. Given that 
$g(r)$ is zero everywhere outside its compact support, these extraneous points do not contribute to the 
integral in Equation \eqref{eqn:convolution_integral_separated_fourier}. This numerical process can be 
repeated for any arbitrary choice of the regularization kernel.

In reality, the force exerted by the particles on the fluid exists for a finite amount of time. Therefore, using 
a singular in time forcing quickly dissipates the generated velocity disturbance. In order to account for the 
finite time contribution of the disturbance, we integrate the velocity contributions from 
Equation \eqref{eqn:convolution_integral_separated_fourier} over time. The resulting net velocity disturbance 
for a constant persistent force applied from $\tau = t_1$ to $\tau = t_2$ is as follows:

\begin{equation}
\mathbf{u}^{\prime}(\mathbf{r},t) = \mathbf{F} \int_{t_1}^{t_2} \left( \int_{\mathbb{R}^{z+}} \mathcal{F}^{-1} \biggl\lbrace \hat{g} (\mathbf{k},z) \star \hat{\mathbf{G}} (\mathbf{k},h,z,\tau) \biggr\rbrace ~\mathrm{d}z \right) \mathrm{d}\tau.
\label{eqn:persistent_convolution_integral_separated_fourier}
\end{equation}

Since the regularization kernel is indepedent of time, Equation \eqref{eqn:persistent_convolution_integral_separated_fourier} 
can be simplified to:

\begin{equation}
\mathbf{u}^{\prime}(\mathbf{r},t) = \mathbf{F} \int_{\mathbb{R}^{z+}} \mathcal{F}^{-1} \biggl\lbrace \hat{g} (\mathbf{k},z) \star \left( \hat{\mathbf{V}} (\mathbf{k},h,z,t_2) - \hat{\mathbf{V}} (\mathbf{k},h,z,t_1)\right) \biggr\rbrace ~\mathrm{d}z,
\label{eqn:persistent_convolution_integral_separated_fourier_antiderivative}
\end{equation}

where,

\begin{equation}
\hat{\mathbf{V}} (\mathbf{k},h,z,t) = \int_{0}^{t} \hat{\mathbf{G}} (\mathbf{k},h,z,\tau) ~\mathrm{d}\tau.
\label{eqn:persistent_stokeslet_fourier}
\end{equation}

$\hat{\mathbf{V}}$ is henceforth referred to as the persistent transient wall-bounded Stokeslet. Equation \eqref{eqn:persistent_convolution_integral_separated_fourier_antiderivative} provides the total 
velocity disturbance contributed by a particle between $t = t_1$ and $t = t_2$ located at a 
distance $h$ away from the wall and which supplies a constant force $\mathbf{F}$ to the fluid in this time 
interval. Since only $\hat{\mathbf{G}}$ is dependent on time in 
Equation \eqref{eqn:persistent_convolution_integral_separated_fourier}, the evaluation of a 
one-dimensional integral is not necessary to calculate the total velocity disturbance. 
Instead, only the difference between the anti-derivatives of $\hat{\mathbf{G}}$ at two times, $t = t_1$ and $t = t_2$ need to be evaluated 
to obtain an approximate velocity disturbance. Expressions for the persistent Stokeslets 
for vertical and horizontal forcing are shown in Section \ref{subsec:appendix_persistentstokeslets_verticalforcing} and \ref{subsec:appendix_persistentstokeslets_horizontalforcing}.

\section{Numerical methodology}
\label{sec:numerical_methodology}

Particles subjected to non-uniform velocities experience forces in arbitrary directions that can vary 
rapidly in time. In such cases, the reaction force from the particle on the fluid changes with 
time and is oriented in arbitrary directions. These reaction forces are responsible for the 
self-induced velocity disturbance of the particle on the fluid. Equation 
\eqref{eqn:persistent_convolution_integral_separated_fourier_antiderivative} provides an expression for this 
velocity disturbance in the fluid for a particle near a planar wall exerting a constant force $\mathbf{F}^{(m)}$ 
on the fluid between the times $t^{(m-1)}$ and $t^{(m)}$. Every subsequent time interval 
results in a new reaction force that carries information about the local nature of the flow around the particle.
In the current model, we account for this changing force contribution from the particles on the fluid by 
using fictitious momentum sources. These fictitious momentum sources are assumed to contribute a constant 
force $\mathbf{F}^{(m)}$ for a given fluid time step, $\Delta t$. Each subsequent time step generates a new 
momentum source that carries information on the magnitude and resultant direction of the force on the fluid. Each 
$\mathbf{F}^{(m)}$ at the $m^{\mathrm{th}}$ time step contributes a velocity disturbance in the fluid given by 
Equation \eqref{eqn:persistent_convolution_integral_separated_fourier_antiderivative}. 
The net velocity disturbance contributed to the fluid after $n$ time steps is as follows:

\begin{equation}
\mathbf{u}_{\mathrm{net}}^{\prime} (\mathbf{x}_{\mathrm{p}}, h, t^{(n)}) = \sum_{m=1}^{n} \mathbf{F}^{(m)} \mathbf{W} (\mathbf{x}_{\mathrm{p}} - \mathbf{x}^{(m)}, h, t^{(m)}).
\label{eqn:netvelocitydisturbance_numericalmethod}
\end{equation}

\begin{figure}[t]
\centering
\includegraphics[scale=1.0]{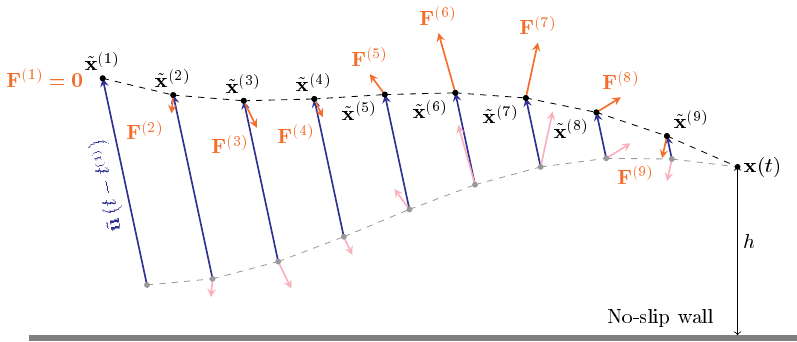}
\caption{Schematic showing the advection of the fictitious momentum sources with the background velocity near a no-slip planar wall, 
generated by a forcing located at $\mathbf{x}_{\mathrm{p}}$. Every fictitious momentum source, $\mathbf{F}^{(m)}$, generated at the $m^{\mathrm{th}}$ time step, 
is advected with the local undisturbed velocity at its center, $\tilde{\mathbf{u}} (\mathbf{x}^{(m)})$. In our simulations, they are advected with the disturbed velocity, $\overline{\mathbf{u}} (\mathbf{x}^{(m)})$. Here, $\mathbf{x}^{(m)}$ 
represents the location of the advected fictitious momentum sources.}
\label{fig:advectingforces_schematic}
\end{figure}

Here $\mathbf{W}$ represents the difference between the pre-computed persistent Stokeslets and is the discrete equivalent of 
the expression in the integral in Equation \eqref{eqn:persistent_convolution_integral_separated_fourier_antiderivative}.
$\mathbf{x}_{\mathrm{p}}$ represents the position of the particle, and $\mathbf{x}^{(m)}$ represents the position of 
the fictitious momentum sources.
Additionally, these generated fluid momentum sources behave like one-way coupled particles and are advected with the 
local undisturbed velocity at the location of these fluid momentum sources, $\tilde{\mathbf{u}} (\mathbf{x}^{(m)})$. 
For a large number of particles in the 
domain, this process becomes intractable, and the evaluation of the undisturbed velocity at the location of these 
momentum sources is a tedious process. Rather, we advect these momentum sources with the local disturbed velocity, 
$\overline{\mathbf{u}} (\mathbf{x}^{(m)})$. 
This is expected to induce an error in the trajectory of these momentum sources but is seen to have an  
insignificant impact on the undisturbed velocity evaluation at the location of the particle, $\tilde{\mathbf{u}} (\mathbf{x}_{\mathrm{p}})$. 
Figure \ref{fig:advectingforces_schematic} shows a schematic of the advection of fictitious momentum sources, 
$\mathbf{F}^{(m)}$ by the background disturbed velocity. Here, forces generated by a particle located at 
$\mathbf{x}_{\mathrm{p}}$ for nine different time instances are shown.

On a computational mesh, the discrete evaluation of the integral in Equation 
\eqref{eqn:persistent_convolution_integral_separated_fourier_antiderivative} is necessary 
to compute $\mathbf{W}$. The force regularization kernel, $g$, and the persistent transient Stokeslet kernel, $\mathbf{V}$, 
are both sampled on a computational grid with spatial resolution $\Delta x$ and temporal resolution $\Delta t$. Here, $\mathbf{V}$ 
is the real space counterpart of the Fourier space Green's function $\hat{\mathbf{V}}$ defined in Section \ref{subsec:regularized_persistent_stokeslet_subsection}. Although in the Appendix 
we derive $\hat{\mathbf{V}}$ in Fourier space, we perform a reverse and forward DFT of $\hat{\mathbf{V}}$ to obtain 
its counterpart $\hat{\mathbf{V}}_{d}$ on a discrete mesh. This process is equivalent to performing an averaging of $\hat{\mathbf{V}}$ on a grid of size 
$\Delta k$, where $\Delta k$ represents the radial wavenumber sampling interval. The reverse DFT of $\hat{\mathbf{V}}$ is performed 
using a Hankel transform as shown in Equation \eqref{eqn:hankel_transform}, and was accelerated using a fast Bessel 
function algorithm \cite{Tumakov2019}. The real-space spatial and temporal sampling intervals $\Delta x$ and $\Delta t$ 
are chosen to match the ones used in the discretization of the fluid governing equations. In the present 
work, we use the second-order space and time accurate, in-house finite volume CFD solver, \texttt{MultiFlow} \cite{Denner2020}.

The temporal sampling of $\mathbf{V}$ 
is done for discrete time levels $t^{(m)} = m \Delta t$, where $m \in [1, K_{\mathrm{max}} ]$. The 
maximum time level $K_{\mathrm{max}}$ is chosen such that the persistent Stokeslet at the point of forcing, 
$\mathbf{x} = \mathbf{r}_0$ attains a steady 
state at $K_{\mathrm{max}} \Delta t$. If $\mathbf{V}_{0} (t)$ represents the persistent Stokeslet at 
$\mathbf{x} = \mathbf{r}_0$, then $K_{\mathrm{max}}$ is determined such that:

\begin{equation}
1 - \frac{\mathbf{V}_0 (t^{K_{\mathrm{max}}})}{\underset{t \to \infty}{\lim} \mathbf{V}_0 (t)} < \epsilon_{\mathrm{tol.}}.
\label{eqn:kmax_infinity_criterion}
\end{equation}

Here $\epsilon_{\mathrm{tol.}}$ represents the prescribed temporal tolerance of the maps, indicating that the discrete maps 
have attained a steady state. Using Equation \eqref{eqn:kmax_infinity_criterion}, 
we determine whether the velocity disturbance field is transient, and Stokeslets beyond $K_{\mathrm{max}}$ can be 
replaced with their steady limit. See Sections \ref{subsec:appendix_persistentstokeslets_verticalforcing} and \ref{subsec:appendix_persistentstokeslets_horizontalforcing} for the $t \to \infty$ limit of $\hat{\mathbf{V}}$.

The maximum velocity contributions from these persistent Stokeslets arise near $\mathbf{x} = \mathbf{r}_0$. 
Additionally, $\mathbf{V}$ is a fast decaying function near the point of impulse, $\mathbf{r}_0$. In order to 
sample $\mathbf{V}$ with second-order spatial accuracy,
we evaluate the central value of the persistent Stokeslets in a cylindrical volume equivalent to the 
volume of the computational cell, $\Delta x^3$. The symmetry of the geometry in the plane parallel to the wall makes using a cylindrical volume 
an ideal choice for analytical integration. For grid points outside this region, the cell-centered 
values of the Stokeslets are used. The strategy can be summarized as follows:

\begin{equation}
\mathbf{V} (\mathbf{x}_i,h,t^{(m)}) = \left\{ \begin{array}{ll} \mathbf{V}_{\mathcal{K}} (\mathbf{x} = \mathbf{r}_0,h,t^{(m)}) & \lVert \mathbf{x}_i \rVert \leq \Delta x /2, \\ \\ \mathbf{V} (\mathbf{x} = \mathbf{x}_i,h,t^{(m)}) & \lVert \mathbf{x}_i \rVert > \Delta x /2. \end{array}\right .
\end{equation}

Here $\mathbf{V}_{\mathcal{K}} (\mathbf{x} = \mathbf{r}_0,h,t^{(m)})$ is the average value of the persistent Stokeslets in a 
cylindrical volume around $\mathbf{x} = \mathbf{r}_0$. Mathematically, this is achieved using a convolution with a 
Top-hat filter defined in cylindrical coordinates. See Section \ref{subsec:centralvalue_regstokeslets} for a detailed derivation. Outside the 
bounds of the mesh cell at $\mathbf{x} = \mathbf{r}_0$, the cell-centered values of the Stokeslets are used, as 
described in Section 
\ref{subsec:appendix_persistentstokeslets_verticalforcing} and 
Section \ref{subsec:appendix_persistentstokeslets_horizontalforcing}.

We utilize the Fourier convolution theorem to perform the convolution of the persistent Stokeslet 
and the discrete regularization kernel in Equation \eqref{eqn:persistent_convolution_integral_separated_fourier_antiderivative}. 
This is defined as follows:

\begin{equation}
\mathbf{V}_{g} (\mathbf{x}_i,h,t^{(m)}) = \sum_{h^{g}_{\mathrm{min}}}^{h^{g}_{\mathrm{max}}} \mathcal{F}^{-1} \lbrace \hat{\mathbf{V}}_{d} (\mathbf{k},h^{g},z,t^{(m)}) \star \hat{g}(\mathbf{k},h^{g})  \rbrace ~\Delta h^{g},
\label{eqn:discrete_convolution_parallelandperpendicular}
\end{equation}

where $h^{g}$ is the wall-normal coordinate signifying the extents $h^{g}_{\mathrm{min}}$ and 
$h^{g}_{\mathrm{max}}$ of the compact regularization kernel whose center lies at $h$, and $\Delta h^{g}$ 
represents to the wall-normal mesh spacing. $\mathbf{V}_{g} (\mathbf{x}_i,h,t^{(m)})$ is the discrete regularized 
transient Stokeslet defined at cell-centered coordinates $\mathbf{x}_i$ of the computational mesh. 
Equation \eqref{eqn:discrete_convolution_parallelandperpendicular} is the discrete equivalent of 
Equation \eqref{eqn:persistent_convolution_integral_separated_fourier_antiderivative}.
The transformed regularization kernel, $\hat{g}(\mathbf{k},h^{g})$, is evaluated using DFT. 
It is to be noted that, for the cases considered in this work, the regularization kernel does not 
overlap with the wall. And therefore, $h^{g}_{\mathrm{min}} > 0$. When the particle is located at $h < \Delta x$ 
from the wall, the radius of the regularization kernel, $\delta = \Delta x / 2$. For the case when the 
regularization kernel radius is larger than the distance of the particle from the wall, i.e., $\delta > h$, 
mirroring the Green's functions $\hat{\mathbf{V}}_{d}$ across the plane representing the wall could be used to 
approximately evaluate the discrete convolution in Equation \eqref{eqn:discrete_convolution_parallelandperpendicular}. 
However, this mirroring does not produce a physical decay of the fluid velocity disturbance, and needs further 
investigation. For the present work, this special case is omitted.

As mentioned in Section \ref{subsec:regularized_persistent_stokeslet_subsection}, since the Green's functions are 
translationally invariant along the wall-normal direction, and in order to make use of the 
varying mesh spacing in the wall-normal direction for wall-bounded flows, we perform the Fourier space 
convolution in the 2D space parallel to the wall and a real space integration in the wall-normal direction. The 
convolution in Fourier space is performed as a point-wise multiplication, making use of the Fourier convolution 
theorem. Using a compact support for the regularization kernel allows the real space convolution in the wall-normal 
direction to be calculated exclusively across the extents of the kernel. The velocity disturbance generated by the 
particle in a Finite-volume solver represents a mesh cell averaged velocity. In order to capture this velocity using a 
semi-analytical approach, the cell-centered regularized Stokeslets are convoluted with a Top-hat filter with support 
$\sqrt[3]{3/4 \pi} \Delta x$ to obtain the volume-averaged Stokeslets, $\overline{\mathbf{V}}_{g} (\mathbf{x}_i,h,t^{(m)})$.

The resulting 
maps are evaluated for various discrete time levels until the steady state time level, $K_{\mathrm{max}}$. The difference between the 
generated maps at time levels $t^{(m-1)}$ and $t^{(m)}$, when multiplied with the corresponding forcing $\mathbf{F}^{(m)}$ 
from the fictitious momentum source, represents the velocity disturbance contribution near a planar wall by a regularized forcing 
in the defined time interval. The net velocity disturbance due to all such fictitious momentum sources is given by 
Equation \eqref{eqn:netvelocitydisturbance_numericalmethod}. 

For a defined regularization kernel and 
distance normal to the wall, $h$, the required velocity disturbance is only a function of the applied forcing 
and the volume-averaged Stokeslet, $\overline{\mathbf{V}}_{g}$. These Stokeslet maps can therefore be pre-computed and stored for 
$K_{\mathrm{max}}$ time levels. For particles moving normal to the wall, these maps 
are evaluated at various wall-normal distances and stored as data files. These are then read as required 
by the particles depending on their wall-normal distance, $h$. For particles moving parallel to the wall, 
$\overline{\mathbf{V}}_{g}$ is translationally invariant 
along the wall-parallel direction, and the resulting maps are unaffected by this shift in real-space parallel to the wall. 
Therefore, it is not necessary to recompute the maps for various positions of the particle in the wall-parallel directions.
Since $\hat{\mathbf{V}}$ is a function of $\nu \Delta t$, where $\nu$ 
is the kinematic viscosity of the fluid, these maps can also be used for computing the velocity disturbance 
at different viscosities by adjusting the time step accordingly such that $\nu \Delta t$ remains a constant.

The velocity disturbance contributed by the momentum source between time levels $t^{(m-1)}$ and $t^{(m)}$ at the center of the real particle depends on the 
relative distance between their centers, $\mathbf{x}_{\mathrm{p}} - \mathbf{x}^{(m)}$. Fictitious momentum sources that have advected far away from the particle 
contribute a velocity disturbance at the location of the particle that is a rapidly decaying function of the 
relative distance. This implies that the regularized maps need to be generated only on a finite spatial stencil. For momentum sources that lie 
outside this stencil but at finite $h$, the Stokeslet maps from the singular Stokeslet can be used. This is because 
the regularized Stokeslet and the singular Stokeslet converge at large $|\mathbf{r}|$.
For momentum sources that are located outside the spatial stencil of the Stokeslet kernel and far away from the wall, 
(i.e., $h \gg 1$), the free-space transient Stokeslet kernel \cite{Pozrikidis1992} can be used. The map generation 
steps at a particular wall-normal distance, $h$ for the transient velocity correction algorithm can be summarized as follows:

\begin{enumerate}[itemsep=-0.5ex]
\item Evaluate Fourier transform, $\hat{g}(\mathbf{k},z)$ of the discrete filter $g(r)$ using DFT.
\item Perform the reverse DFT of $\hat{\mathbf{V}}$. Reverse transform to real space using Equation \eqref{eqn:hankel_transform} and obtain 
$\mathbf{V} (\mathbf{x}_i,h,t^{(m)})$.
\item Compute cell-centered persistent Stokeslets for $\lvert \mathbf{r} \rvert > \Delta x /2$ using Section \ref{subsec:appendix_persistentstokeslets_verticalforcing} and 
Section \ref{subsec:appendix_persistentstokeslets_horizontalforcing}, and for $\lvert \mathbf{r} \rvert \leq \Delta x / 2$ using Section \ref{subsec:centralvalue_regstokeslets}.
\item Evaluate the convolution of the persistent Stokeslets and DFT to real space using Equation \eqref{eqn:discrete_convolution_parallelandperpendicular} to obtain $\mathbf{V}_{g} (\mathbf{x}_i,h,t^{(m)})$.
\item Obtain the volume-averaged Stokeslet kernel, $\overline{\mathbf{V}}_{g} (\mathbf{x}_i,h,t^{(m)})$, via convolution with a Top-hat filter with support $\sqrt[3]{3/4 \pi} \Delta x$.
\item Store the resulting maps into an array for various wall-normal distances, $h$, which can be read later from files.
\item Repeat steps 1 to 6 for various time levels, $t^{(m)}$, until $m = K_{\mathrm{max}}$, where $K_{\mathrm{max}}$ is defined by Equation \eqref{eqn:kmax_infinity_criterion}.
\end{enumerate}

This process can be repeated for various $h$ to generate the corresponding maps. 
The relation between $\overline{\mathbf{V}}_g$ and $\mathbf{W}$ (in Equation \eqref{eqn:netvelocitydisturbance_numericalmethod}) can now be expressed as,

\begin{equation}
\mathbf{W} (\mathbf{x}_{\mathrm{p}} - \mathbf{x}^{(m)}, h, t^{(m)}) = \overline{\mathbf{V}}_{g} (\mathbf{x}_{\mathrm{p}} - \mathbf{x}^{(m)},h,t^{(m)}) - \overline{\mathbf{V}}_{g} (\mathbf{x}_{\mathrm{p}} - \mathbf{x}^{(m)},h,t^{(m-1)}).
\label{eqn:regularized_persistent_stokeslet_antiderivative}
\end{equation}

The computed fluid velocity disturbance, using Equation \eqref{eqn:netvelocitydisturbance_numericalmethod} and \eqref{eqn:regularized_persistent_stokeslet_antiderivative}, 
is interpolated to the particle center using a trilinear interpolation scheme. The resulting 
undisturbed velocity at the particle center is obtained by subtracting this interpolated disturbance velocity from 
the interpolated disturbed velocity as in Equation \eqref{eqn:disturbed_undist_velocity_relationship}.

\begin{figure}[H]
\includegraphics[scale=1.0]{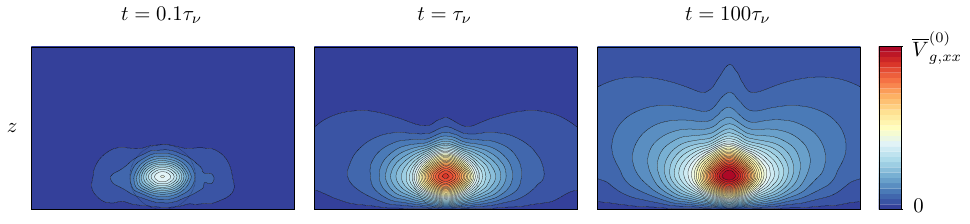}
\includegraphics[scale=1.0]{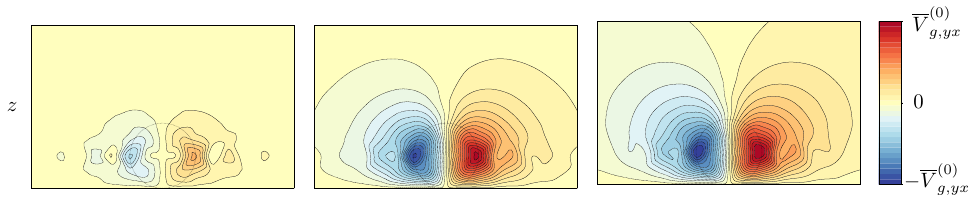}
\includegraphics[scale=1.0]{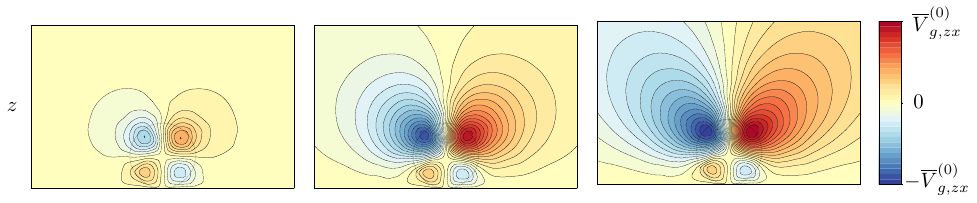}
\includegraphics[scale=1.0]{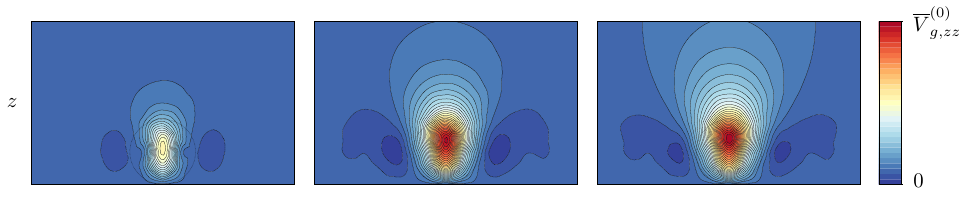}
\includegraphics[scale=1.0]{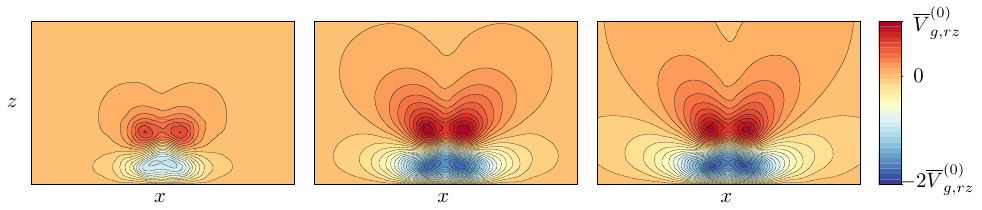}
\caption{Maps showing the regularized persistent transient Stokeslet kernels near a planar wall, $\overline{\mathbf{V}}_{g} (\mathbf{x}_i,h,t)$ 
at a wall-normal distance $h/\delta = 1$ away from the wall, in the $y = \mathbf{r}_0 \cdot \mathbf{e}_{y}$ plane, and at three different times of $t = \{0.1 \tau_{\nu}, \tau_{\nu}, 100 \tau_{\nu} \}$. 
Here $\tau_{\nu} = a^2/\nu$ represents the viscous dissipation time scale. $\overline{V}_{g,xx}$, $\overline{V}_{g,yx}$, 
and $\overline{V}_{g,zx}$ represent the $x$, $y$, and $z$ velocity responses respectively for a forcing parallel to the wall (located at $z = 0$) along the $x$ direction. 
$\overline{V}_{g,zz}$ and $\overline{V}_{g,rz}$ represent the $z$, and $r$ velocity responses respectively for a forcing perpendicular to the 
wall along the $z$ direction. All plots are normalized with their corresponding maximum steady state ($t \to \infty$) values, $\overline{\mathbf{V}}_{g}^{(0)}$. The dotted black circle represents the region of regularization with a radius $\delta$.}
\label{fig:regularizedkernel_allmaps}
\end{figure}

Figure \ref{fig:regularizedkernel_allmaps} shows the generated regularized persistent transient Stokeslet kernels $\overline{\mathbf{V}}_{g} (\mathbf{x}_i,h,t)$ 
near a planar wall at a wall-normal distance $h/\delta = 1$ away from the wall. The maps are depicted for three different normalized times: 
$t = \{0.1 \tau_{\nu}, \tau_{\nu}, 100 \tau_{\nu} \}$, where $\tau_{\nu} = a^2/\nu$ is the viscous dissipation time scale. Here, $a$ is defined as follows \cite{Anderson1967,Evrard2024a}:

\begin{equation}
4 \pi \int_{0}^{a} g(r) r^2 ~\mathrm{d}r = \frac{1}{2}. 
\end{equation}

For a Wendland kernel, $a \approx 0.4107 \delta$. The dotted circle represents the region of regularization with a kernel 
of radius $\delta$. The maps depict the velocity responses in the $x$, $y$, and $z$ 
directions as $\overline{V}_{g,xx}$, $\overline{V}_{g,yx}$, and $\overline{V}_{g,zx}$ respectively 
for a forcing in the horizontal direction ($x$) parallel to the wall. Similarly, the velocity responses in the $z$ and $r = \sqrt{x^2 + y^2}$ (radial)
directions are denoted by $\overline{V}_{g,zz}$ and $\overline{V}_{g,rz}$ respectively for a forcing 
in the vertical direction ($z$) perpendicular to the wall. The decay of the kernels to zero at $z = 0$ to satisfy the no-slip condition 
at the wall is observed for all cases. The $\overline{V}_{g,xx}$ component produces a symmetric field around the 
point of forcing. Comparatively, the transverse and wall-normal components, $\overline{V}_{g,yx}$ and $\overline{V}_{g,zx}$, 
produce an anti-symmetric field around $\mathbf{r}_0$. A similar trend can be observed for the vertical forcing, where the 
$\overline{V}_{g,zz}$ component is symmetric, while the $\overline{V}_{g,rz}$ component is anti-symmetric. This is a consequence 
of the divergence-free condition of the Green's functions.

The discrete convolution of the transient Stokeslet kernel with the regularization kernel, $g(r)$, offers the advantage of using 
arbitrary regularization kernels. The use of horizontal and vertical Stokeslet maps allows the estimation of the velocity disturbance 
when $\mathbf{F}^{(m)}$ is oriented in arbitrary directions. The linearity of the velocity disturbance 
governing equations enables the superposition of the disturbance contribution from each of these directions, as described in 
Section \ref{subsec:velocity_disturbance_governing_equations}. Additionally, since the model depends on the analytical 
Stokeslet kernel, it does not require any fitted parameters to enforce the no-slip condition at the wall. 
The pre-computation of the maps makes it a one-time cost to simulate any number of particles in the domain near a planar wall.

\section{Results}
\label{sec:results}

\quad In this section, we test the applicability of our current model on both stationary and moving particles. 
All simulations are performed on a computational mesh of uniform isotropic spacing. As the first case, we 
analyze the mesh convergence of our model by studying the recovery of the undisturbed velocity for a fixed particle 
near a planar wall subjected to a uniform flow. This test case helps to identify the $d_{\mathrm{p}}/\Delta x$ regimes under 
which a velocity disturbance correction becomes necessary. We then examine cases involving a single particle
moving parallel to the wall and a single particle free-falling under an external acceleration perpendicular to the planar wall. 
The moving particle cases are studied on cases ranging from low to moderate Stokes numbers. Unless stated otherwise, 
a regularization kernel radius to mesh spacing ratio, $\delta/ \Delta x = 1$ is used.

\subsection{\textnormal{\textit{Fixed particle near a planar wall in a uniform flow}}}

We investigate the effect of a finite-mesh resolution on the wall-bounded fluid velocity correction through a 
test case involving a fixed particle subjected to a uniform flow at a distance $h$ from a planar no-slip wall.
The undisturbed fluid velocity at the particle center in such a case is obtained from the corresponding 
one-way coupled EL simulation. Our objective is to identify if the momentum 
diffusion from the particle on a computational mesh is captured using our semi-analytical velocity disturbance model. 
To do this, we study the problem 
in a fully-developed steady velocity field, decoupling the transient nature of the boundary layer development over 
the planar wall. The advection of the particle-generated velocity disturbance on the fluid 
by the developing boundary layer is not captured by the current model. Although the transient 
fluid-particle momentum sources are advected with the background disturbed velocity (ideally, the undisturbed 
velocity), the boundary layer 
development being a nonlinear phenomena means that a linear advection model for the fluid-particle momentum sources 
cannot capture this effect.

Numerical simulations have studied the drag on a spherical particle near a planar wall in shear flow, 
and a correlation has been obtained as a function of the particle separation from the wall 
and the local Reynolds number \cite{Zeng2009,Lee2010a}. 
At distances far from the wall, i.e., as $h \to \infty$, this correlation tends to the 
well-known Schiller-Naumann correlation \cite{Schiller1933}. For the fixed-particle setup, 
we use the following drag correlation \cite{Zeng2009}:

\begin{equation}
\mathrm{C}_{\mathrm{D}} = \frac{24}{\mathrm{Re}_{\mathrm{p}}} \Biggl( 1 + 0.138 \exp(-2l) + \frac{9}{16 (1 + 2l)} \Biggr) (1 + \alpha_s \mathrm{Re}_{\mathrm{p}}^{\beta_s}),
\label{eqn:dragcorrelation_zeng2009}
\end{equation}

where $l = h/d_{\mathrm{p}} - 0.5$ is the distance between the wall and the closest point on the sphere to the wall, 
$\alpha_s = 0.15 - 0.046 (1 - 0.16 l^2) \exp(-0.7l)$, and $\beta_s = 0.687 + 0.066 (1 - 0.76 l^2) \exp(-l^{0.9})$.
Due to the decreasing drag coefficient for the particle with increasing Reynolds number ($\mathrm{Re}_{\mathrm{p}}$), the magnitude 
of the velocity disturbance decreases with $\mathrm{Re}_{\mathrm{p}}$, reducing the deviation of 
$\tilde{\mathbf{u}}$ from $\overline{\mathbf{u}}$. Therefore, we test our model at a low Reynolds number of $10^{-3}$. 
At higher $\mathrm{Re}_{\mathrm{p}}$, in a steady laminar boundary layer, the importance of a velocity correction 
model is less significant \cite{Evrard2021}.

We test our current model for a particle with varying particle diameter to mesh-spacing ratios $d_{\mathrm p}/\Delta x$, 
at various heights $h/\delta$ away from the wall, ensuring that the local Reynolds number at the particle center based on 
the reference undisturbed velocity is $10^{-3}$. A nondimensional length scale of $h/d_{\mathrm{p}}$ might seem like a natural choice 
to characterize the fluid velocity disturbance. 
However, the velocity disturbance generated due to a regularized forcing scales with the regularization length scale, $\delta$. 
Therefore, $\delta$ represents a more physical choice for the characteristic length scale that determines the length scale of momentum diffusion due to the forcing. 
We used a mesh resolution of 80 cells across the wall-normal direction, which is found to be sufficient for producing grid-independent results.

\begin{figure}[t]
\centering
\includegraphics[scale=1.0]{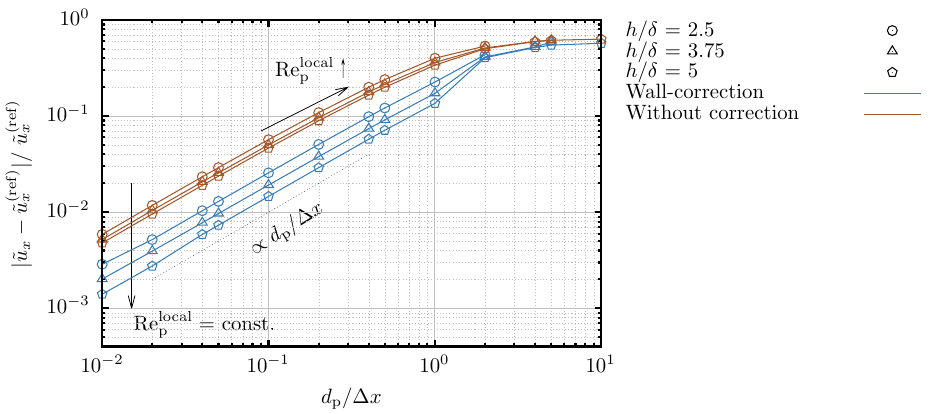}
\caption{Relative error in the streamwise undisturbed velocity estimation for a fixed particle located at various distance $h$ 
away from a planar wall subjected to a uniform flow. The particle diameter to mesh-spacing ratio is varied from 
$d_{\mathrm p}/\Delta x = 0.01$ to $d_{\mathrm p}/\Delta x = 10$. $h$ is varied for each $d_{\mathrm p}/\Delta x$ such that 
the local particle Reynolds number $\mathrm{Re}_{\mathrm{p}}^{\mathrm{local}} = \tilde{u}_x (\mathbf{x}_{\mathrm{p}}) d_{\mathrm{p}} / \nu_{\mathrm{f}}$ remains constant.}
\label{fig:fixedparticle_relativeerror}
\end{figure}

Figure \ref{fig:fixedparticle_relativeerror} shows the relative error of the undisturbed velocity obtained 
with and without a wall-based correction to the reference undisturbed velocity from a one-way coupled EL 
simulation. We explore a parameter space ranging from $d_{\mathrm p}/\Delta x = 0.01$ to $d_{\mathrm p}/\Delta x = 10$ 
for three different normalized distances $h/\delta$ from the wall. The fluid flow remains the same in the absence of 
the particle, resulting in a constant boundary layer thickness for each $d_{\mathrm{p}}/\Delta x$ 
within each $h/\delta$. Here, $\overline{u}_x^{\mathrm{ref}}$ represents the undisturbed velocity obtained from 
a one-way coupled simulation at the particle center and serves as the reference solution for this case. The particle 
is positioned at a distance of $h = 0.7 \delta_{99}$ for each case, where $\delta_{99}$ is the thickness of the 
Blasius laminar boundary layer near a planar wall, ensuring a local Reynolds number of 
$\mathrm{Re}_{\mathrm{p}}^{\mathrm{local}} = \tilde{u}_x (\mathbf{x}_{\mathrm{p}}) d_{\mathrm{p}} / \nu_{\mathrm{f}} = 10^{-3}$. 
The scaling of the relative velocity error is observed to 
increase linearly with the particle diameter to mesh-spacing ratio for both the uncorrected and wall-based corrected 
undisturbed velocities at every $h/\delta$. The error in the undisturbed velocity is at least 
three times larger without correction compared to the cases with the wall-bounded correction. This error also increases with decreasing $h/\delta$, 
both with and without the correction. For a particle located very close to the wall, 
although the magnitude of the self-induced velocity disturbance is small, an anisotropic mesh resolution is necessary to 
capture its decay near the wall.
As the particle size increases with increasing $d_{\mathrm p}/\Delta x$, 
$\mathrm{Re}_{\mathrm{p}}^{\mathrm{local}}$ increases, reducing the importance of the velocity correction, 
and the difference between the disturbed and undisturbed velocities decreases. However, the error remains 
large due to the large size of the particle compared to the computational mesh spacing.

\subsection{\textnormal{\textit{Particle settling parallel to a planar wall in a quiescent flow}}}

\quad We also validate our model for a rigid spherical particle falling under gravity parallel to the no-slip wall in 
a quiescent flow. The particle velocity increases until it eventually reaches a terminal velocity when the 
weight of the particle balances with buoyancy and drag forces. In a two-way coupled simulation, the self-induced 
velocity disturbance caused by the particle in the fluid results in a decrease in the predicted drag, 
which eventually leads to a higher terminal velocity of the particle. Hence, the reference 
solution shown in this section for all cases is obtained using a one-way coupled simulation. 
The nondimensional parameters of interest for moving particles is the Stokes number, 
$\mathrm{St} = \tau_{\mathrm{p}}/\tau_\mathrm{f}$. Here $\tau_{\mathrm{p}}$ is the particle time scale of motion
defined as follows:

\begin{equation}
\tau_{\mathrm{p}} = \frac{\rho_{\mathrm{p}} d^2_{\mathrm{p}}}{18 \mu_{\mathrm{f}}},
\label{eqn:particle_timescale}
\end{equation}

and $\tau_{\mathrm{f}}$ is the appropriate fluid time scale. For the cases discussed here, 
we remain in the diffusion-dominated regime, and hence,

\begin{equation}
\tau_{\mathrm{f}} = \frac{\Delta x^2}{\nu}.
\label{eqn:fluid_timescale}
\end{equation}

When $\mathrm{St} \gg 1$, the fluid time scale is smaller than the particle response time scale. In such cases, the velocity disturbance by the particle 
within the fluid has attained a steady state within $\tau_{\mathrm{p}}$, and using a steady disturbance model 
is sufficient to correct for this velocity disturbance \cite{Evrard2020a}. On the other hand, when the particle time scale is smaller 
compared to the fluid time scale, 
i.e., $\mathrm{St} \ll 1$, the velocity disturbance in the fluid is transient, and an unsteady disturbance 
model is required to obtain the undisturbed velocity. Using a steady disturbance model for a particle when $\mathrm{St} \ll 1$ results in 
overprediction of the disturbance and an underprediction of the terminal velocity. 
Another nondimensional parameter of interest is the particle Reynolds number,

\begin{equation}
\mathrm{Re^{(fs)}_p} = \frac{\lvert \mathbf{u}^{(\mathrm{fs})} \rvert d_{\mathrm{p}}}{\nu_{\mathrm{f}}},
\label{eqn:reynoldsnumber_settlingparticle_definition}
\end{equation}

where

\begin{equation}
\mathbf{u}^{(\mathrm{fs})} = \frac{(\rho_{\mathrm{p}} - \rho_{\mathrm{f}})}{\rho_{\mathrm{p}}} \tau_{\mathrm{p}} \mathbf{g}
\end{equation}

represents the particle settling velocity in an unbounded domain when subjected to an acceleration, $\mathbf{g}$. Particles near wall 
are subjected to an increased shear. \citet{Zeng2009} provide drag correlations for a particle moving parallel to a wall in a 
quiescent flow:

\begin{equation}
\frac{\mathrm{C_{\mathrm{D}\textit{t}}}}{\mathrm{C_{\mathrm{D}\textit{t0}}}} = 1 + \alpha_t \mathrm{Re_p}^{\beta_t},
\label{eqn:zeng_particlesettlingparallel_dragcorrelation}
\end{equation}

where

\begin{equation}
\mathrm{C_{\mathrm{D}\textit{t0}}} = \frac{24}{\mathrm{Re_p}} \Biggl[1.028 - \frac{0.07}{1 + 4l^2} - \frac{8}{15} \log \left( \frac{270 l}{135 + 256 l} \right) \Biggr].
\end{equation}

Here $\mathrm{C_{\mathrm{D}\textit{t0}}}$ is the low Reynolds number limit of the correlation, $\alpha_t = 0.15 [1 - \exp(-\sqrt{l})] $, 
and $\beta_t = 0.687 + 0.313 \exp(-2 \sqrt{l})$ are power law factors that are a function of the separation between the particle and the wall. 
In the limit of $l \gg 1$, these correlations simplify to the standard Schiller-Naumann correlation \cite{Schiller1933}.
Similarly, the lift correlations used were as follows \cite{Zeng2009}:

\begin{equation}
\mathrm{C}_{\mathrm{L\textit{t}}} = \mathrm{C}_{\mathrm{L\textit{t0}}} a_{\mathrm{L\textit{t}}}^2 \left( \frac{L}{1.5} \right)^{-2 \tanh(0.01 \mathrm{Re_p})},
\label{eqn:zeng_settlingparallel_liftcorrelation}
\end{equation}

where

\begin{equation}
\mathrm{C}_{\mathrm{L\textit{t0}}} = \left\{ \begin{array}{ll} (9/8 + 5.78 \times 10^{-6} L^{\star 4.58}) \exp(-0.292 L^{\star}) & \mathrm{for} ~L^{\star} \in (0,10) \\ 8.94 L^{\star -2.09} & \mathrm{for} ~L^{\star} \in (10,300) \end{array}\right . \
\end{equation}

Here, $L^{\star} = \mathrm{Re_p} h/d_{\mathrm{p}}$ is the separation between the particle center and the wall, scaled by the local Reynolds number. $\mathrm{C}_{\mathrm{L\textit{t0}}}$ is the experimental low Reynolds number lift correlation for a spherical particle \cite{Vasseur1977}. 
The power law factor correlating the lift for various $\mathrm{Re_p}$ is 
$a_{\mathrm{L\textit{t}}} = 1 + 0.6 \mathrm{Re_p}^{1/2} - 0.55 \mathrm{Re_p}^{0.08}$ \cite{Takemura2003}.

Particles were allowed to settle parallel to the wall at various normalized wall-normal distances $h/\delta$ for 
particle-diameter to mesh-spacing ratios of $d_{\mathrm{p}} / \Delta x = \{0.1, 0.5, 1 \}$. 
All cases are performed at an unbounded particle Reynolds number, $\mathrm{Re^{(fs)}_p} = 0.01$ 
on a uniform mesh-spacing. 
At such low $\mathrm{Re^{(fs)}_p}$, the particle experiences relatively large drag forces and the momentum diffusion 
length scale is large, making it an ideal parameter space where velocity disturbance corrections 
become important for accurately predicting the undisturbed fluid velocity. At high $\mathrm{Re^{(fs)}_p}$, the deviation of the disturbed from the undisturbed 
velocity is expected to decrease owing to the lower drag and smaller momentum diffusion around the particle.

\begin{figure}[H]
\centering
\includegraphics[scale=1.0]{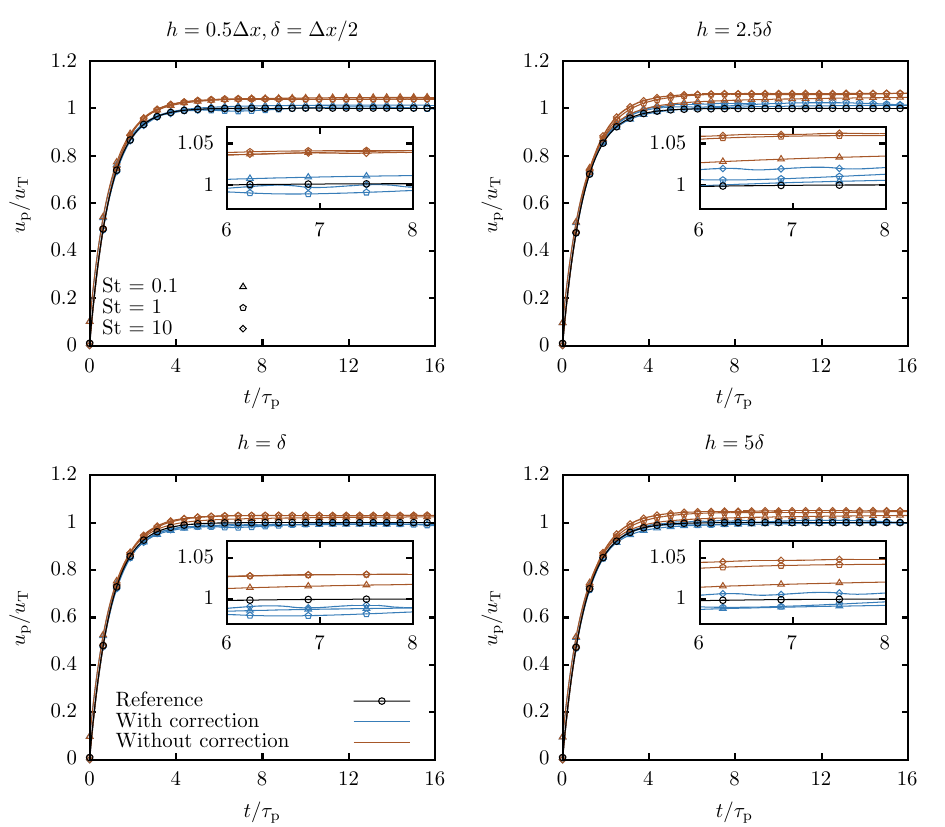}
\caption{Settling velocity of a particle moving parallel to a planar wall at $d_{\mathrm{p}}/\Delta x = 0.1$, normalized by the 
theoretical settling velocity $u_{\mathrm{T}}$ from a one-way coupled EL simulation. The particle settling is compared at different wall-normal 
distances $h/\delta = \{0.5, 1, 2.5, 5 \}$. The colors differentiate the uncorrected and corrected simulations and the 
markers indicate the three different Stokes numbers used, $\mathrm{St} = \{0.1, 1, 10 \}$.
All cases are performed at an unbounded Reynolds number, $\mathrm{Re_p^{(fs)}} = 0.01$. The inset plots 
provide a detailed view of $u_{\mathrm{p}}/u_{\mathrm{T}}$ between $t/\tau_{\mathrm{p}} = 6$ and $t/\tau_{\mathrm{p}} = 8$.}
\label{fig:settlingparticle_parallel_dpbydx0p1}
\end{figure}

\begin{figure}[H]
\centering
\includegraphics[scale=1.0]{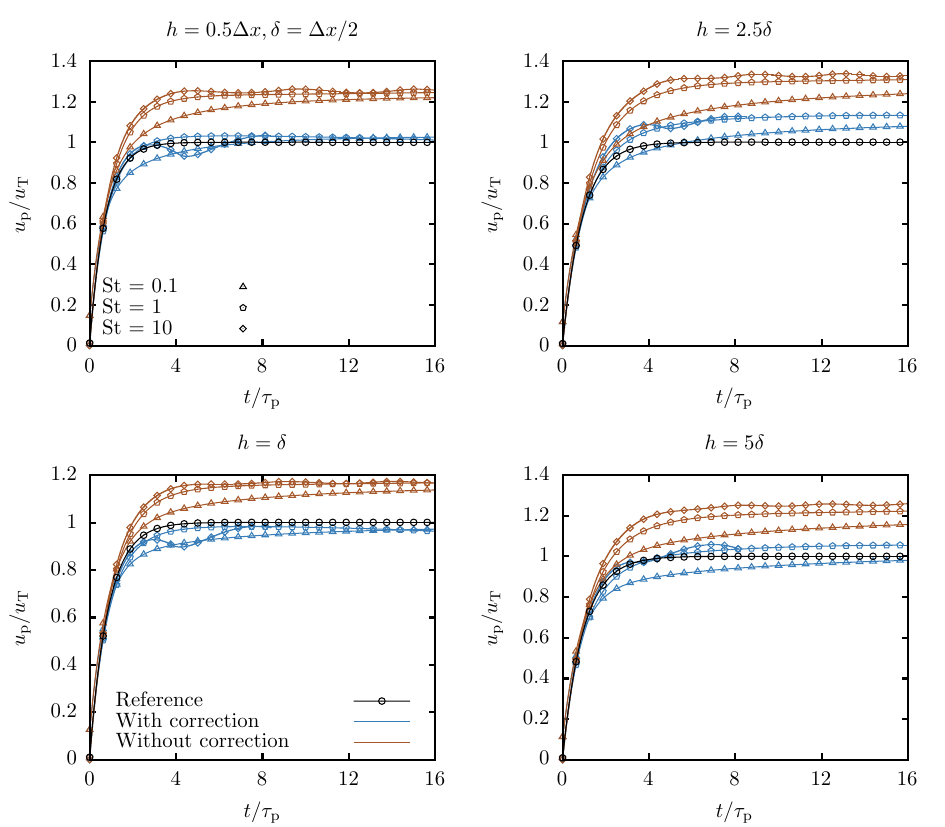}
\caption{Settling velocity of a particle moving parallel to a planar wall at $d_{\mathrm{p}}/\Delta x = 0.5$, normalized by the 
theoretical settling velocity $u_{\mathrm{T}}$ from a one-way coupled EL simulation. The particle settling is compared at different wall-normal 
distances $h/\delta = \{0.5, 1, 2.5, 5 \}$. The colors differentiate the uncorrected and corrected simulations and the 
markers indicate the three different Stokes numbers used, $\mathrm{St} = \{0.1, 1, 10 \}$.
All cases are performed at an unbounded Reynolds number, $\mathrm{Re_p^{(fs)}} = 0.01$.}
\label{fig:settlingparticle_parallel_dpbydx0p5}
\end{figure}

\begin{figure}[H]
\centering
\includegraphics[scale=1.0]{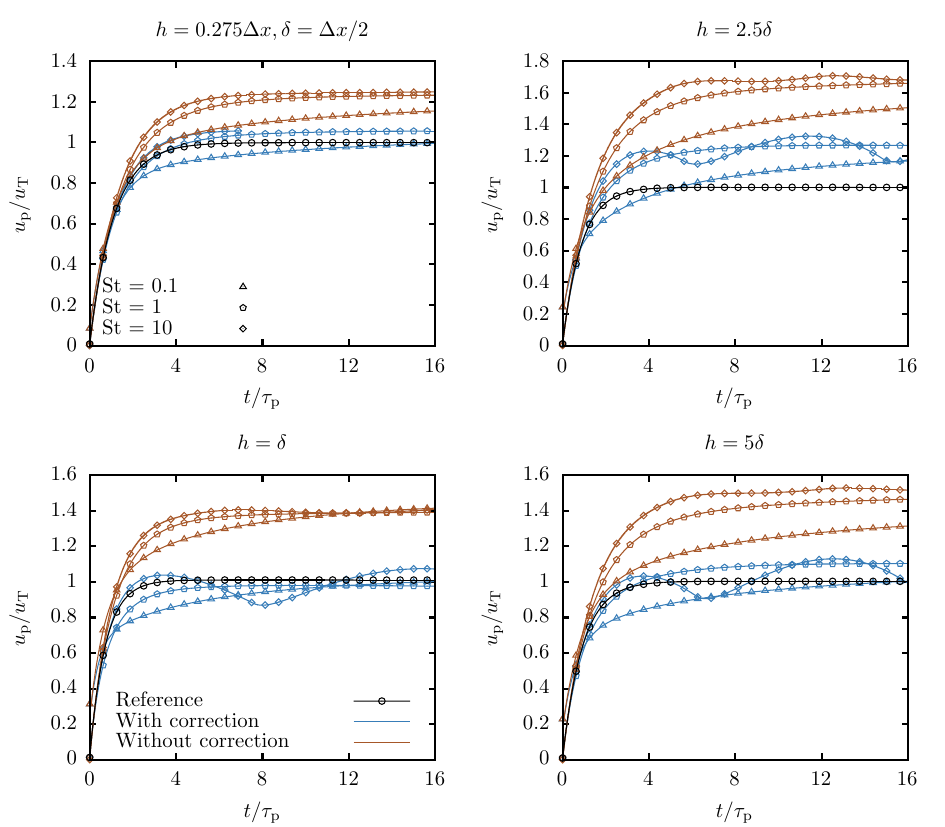}
\caption{Settling velocity of a particle moving parallel to a planar wall at $d_{\mathrm{p}}/\Delta x = 1$, normalized by the 
theoretical settling velocity $u_{\mathrm{T}}$ from a one-way coupled EL simulation. The particle settling is compared at different wall-normal 
distances $h/\delta = \{0.275, 1, 2.5, 5 \}$. The colors differentiate the uncorrected and corrected simulations and the 
markers indicate the three different Stokes numbers used, $\mathrm{St} = \{0.1, 1, 10 \}$.
All cases are performed at an unbounded Reynolds number, $\mathrm{Re_p^{(fs)}} = 0.01$.}
\label{fig:settlingparticle_parallel_dpbydx1}
\end{figure}

Figures \ref{fig:settlingparticle_parallel_dpbydx0p1}, \ref{fig:settlingparticle_parallel_dpbydx0p5}, and 
\ref{fig:settlingparticle_parallel_dpbydx1} depict the normalized settling velocity of a particle moving parallel to a wall 
as a function of the time $t$ normalized by the particle time scale, $\tau_\mathrm{p}$. The settling velocity 
of the particle is normalized with the terminal velocity from a corresponding one-way coupled EL simulation, which  
serves as the reference solution in the figures. Simulations are conducted 
at three different Stokes numbers, $\mathrm{St} = \{0.1, 1, 10 \}$. In all cases, the 
predicted settling velocity error is maximum for $h/\delta = 2.5$. This is a 
consequence of using a uniform mesh-spacing near the wall, and a finite-volume solver does not fully resolve the 
velocity disturbance with this resolution at a low Reynolds number of $\mathrm{Re_p^{(fs)}} = 0.01$. Similar oscillations in the mesh-generated velocity disturbance with respect to $h$, without any velocity correction is also observed in \citet{Pakseresht2020a}. This artifact is 
reflected in both the corrected and uncorrected cases, and therefore does not affect the correction procedure itself. 
For all cases, simulations with 
the wall-bounded correction results in a significantly lower error compared to simulations without correction. 
Furthermore, the error in the settling velocity is observed to increase with a higher particle-diameter to 
mesh-spacing ratio, $d_{\mathrm{p}}/\Delta x$, for both corrected and uncorrected simulations. 
Even for the case with $d_{\mathrm{p}}/\Delta x = 0.1$, where the velocity disturbance of the particle on the 
Eulerian mesh is small, the maximum deviation of the particle settling velocity from the reference solution is 
6.31 \% for the uncorrected simulation at $\mathrm{St} = 10$. Using the wall-bounded correction, this error is reduced to 1.9 \%.

Similarly, for $d_{\mathrm{p}}/\Delta x = 0.5$ and $1$, the maximum errors for the uncorrected simulations are 33.28 \% and 
69.42 \%, respectively. In contrast, the errors for the corresponding cases using the wall-bounded correction are 9.81 \% and 17.08 \% 
respectively. The largest deviation in the particle settling velocity in all cases is observed for 
$\mathrm{St} = 10$. This is due to the relatively larger particle inertia compared to the Eulerian fluid cell it resides in, 
resulting in the particle contributing a larger magnitude of the velocity disturbance to the Eulerian fluid cell.
The oscillations observed at all Stokes numbers is due to the use of a trilinear interpolation scheme to interpolate 
the disturbed fluid velocity and the velocity disturbance from the model to the particle center. This effect is more pronounced 
for the case of $\mathrm{St} = 10$, where the 
particle inertia is large, leading to a particle changing interpolation stencils much more rapidly compared to 
lower Stokes numbers. It is to be noted that for the case of $h = 0.275 \Delta x$, 
the separation of the particle from the no-slip wall is, $l = 0.05$.

\begin{figure}[H]
\centering
\includegraphics[scale=1.0]{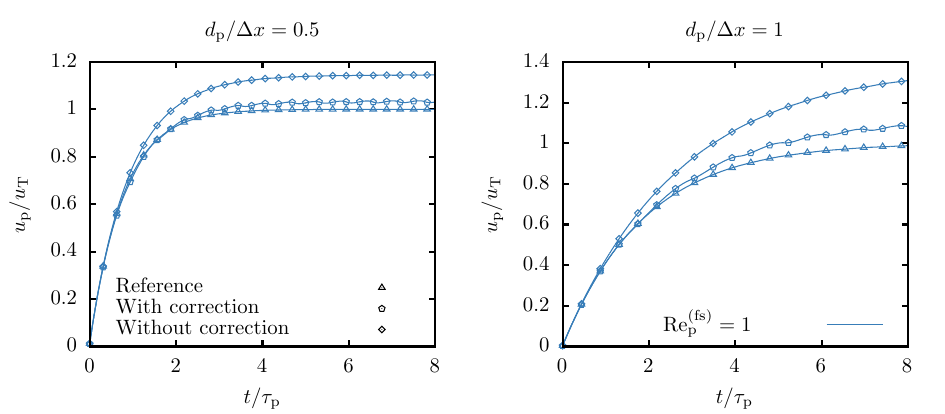}
\caption{Settling velocity of a particle moving parallel to a planar wall at $h/\delta = 1$, normalized by the 
theoretical settling velocity $u_{\mathrm{T}}$ from a one-way coupled EL simulation. The particle settling is compared at different 
particle diameter to mesh-spacing ratios, $d_{\mathrm{p}}/\Delta x = \{0.5, 1 \}$. The markers differentiate the uncorrected, wall-corrected, and 
reference simulations. All cases are performed at an unbounded Reynolds number $\mathrm{Re_p^{(fs)}} = 1$, 
Stokes number $\mathrm{St} = 1$, and $\delta/\Delta x = 1$.}
\label{fig:settlingparticle_parallel_redependence}
\end{figure}

Figure \ref{fig:settlingparticle_parallel_redependence} compares the normalized settling velocity for $\mathrm{Re_p^{(fs)}} = 1$ 
and $\mathrm{St} = 1$ between using the correction model and without any correction. The simulations which do not use a model for the fluid disturbance 
velocity show an error of 14.39 \% in the particle settling velocity estimation, 
whereas when the model for the wall-bounded correction is used, this error reduces to 3.33 \% for the $d_{\mathrm{p}}/\Delta x = 0.5$ case.

\subsection{\textnormal{\textit{Free-falling particle in a quiescent flow perpendicular to a planar wall}}}

Particles near walls additionally experience wall-normal motion, and recovering the correct undisturbed velocity in such 
configurations has 
recently received attention \cite{Pakseresht2020a,Pakseresht2021}.
In this section, we examine the wall-normal motion of the particle using the test case of a free-falling particle 
under acceleration normal to the no-slip wall. The particle slows down during its motion due to lubrication forces acting 
on it. In the point particle framework, this lubrication effect is accounted for using 
the following drag correlation:

\begin{equation}
\mathrm{C_{\mathrm{D}}}^{\perp} = \frac{24}{\mathrm{Re_p}} f(l),
\end{equation}

where

\begin{equation}
f(l) = \left\{ \begin{array}{ll} 1 + \left( \frac{0.562}{1 + 2l} \right) & l > 1.38 ~~\textnormal{\cite{Brenner1961}} , \\ \\ \frac{1}{2l} \left(1 + 0.4 l \log(\frac{1}{2l}) + 1.94 l \right) & l < 1.38 ~~\textnormal{\cite{Cox1967}}. \end{array}\right .
\end{equation}

Here $f(l)$ corrects the unbounded drag coefficient for the separation of the particle from the 
planar wall. 

\begin{figure}[H]
\centering
\includegraphics[scale=1.0]{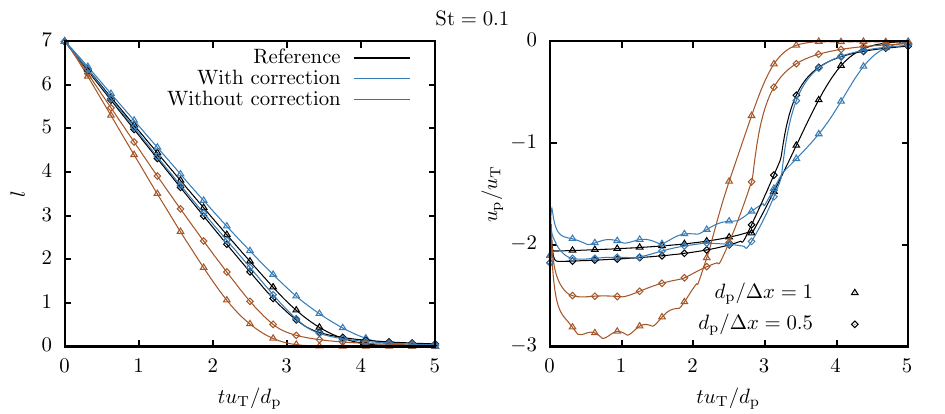}
\vspace{-2mm}
\includegraphics[scale=1.0]{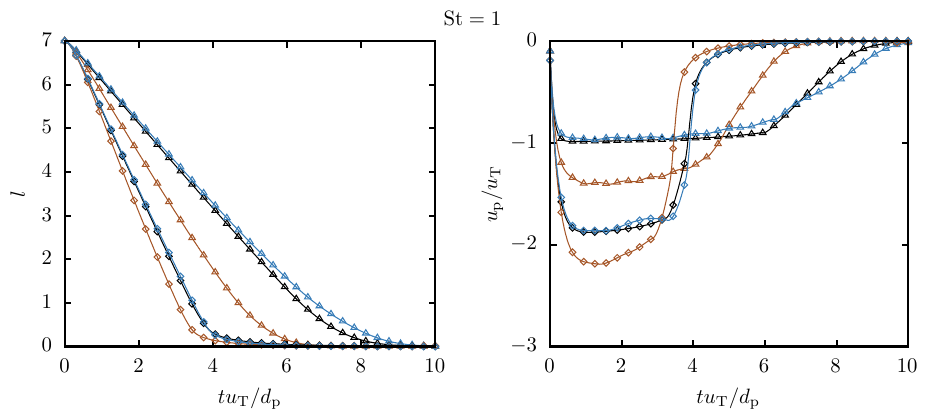}
\vspace{-2mm}
\includegraphics[scale=1.0]{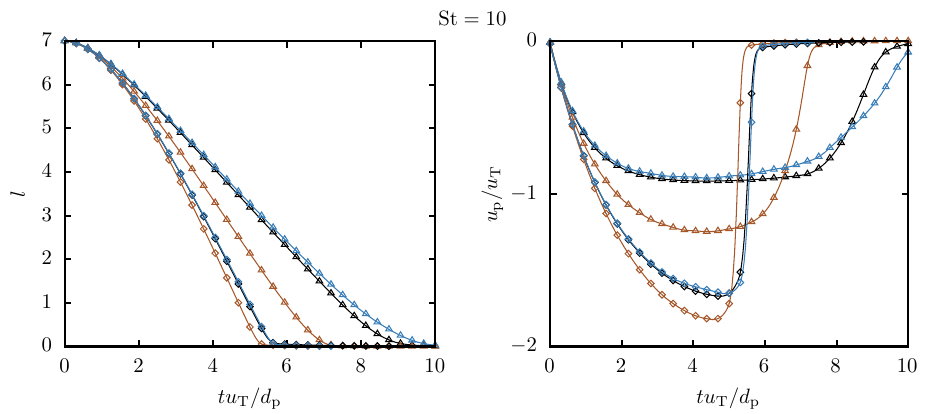}
\caption{Normalized separation, $l = h/d_{\mathrm{p}} - 0.5$ and velocity, $u_{\mathrm{p}}/u_{\mathrm{T}}$ of a particle falling perpendicular to a planar wall at three different Stokes numbers, 
$\mathrm{St} = \{0.1, 1, 10 \}$. The velocity is normalized by the settling velocity $u_{\mathrm{T}}$ in an unbounded scenario. 
The particle settling is compared at two different particle diameter to mesh-spacing ratios, $d_{\mathrm{p}}/\Delta x = \{0.5, 1 \}$. 
The colors differentiate the uncorrected, wall-corrected, and reference simulations. All cases are performed at an unbounded Reynolds number, 
$\mathrm{Re_p^{(fs)}} = 0.1$, and $\delta/\Delta x = 1$.}
\label{fig:settlingparticle_perpendicular}
\end{figure}

As a test case, we subject an initially stationary particle to a constant external acceleration in the wall-normal direction, initially positioned at a normalized separation of $l = 7$.
Figure \ref{fig:settlingparticle_perpendicular} compares the normalized distance ($l = h/d_{\mathrm{p}} - 0.5$) and normalized settling velocity ($u_{\mathrm{p}}/u_{\mathrm{T}}$)
of a particle using a wall-bounded correction against the corresponding reference and simulations without any velocity disturbance correction. The normalization velocity, $u_{\mathrm{T}}$ 
is calculated for a particle settling in an unbounded flow in the Stokes regime. The reference solutions are obtained via a one-way coupled 
EL simulation, where the undisturbed velocity at the particle center is exactly zero. Simulations are performed for two different particle diameter 
to mesh-spacing ratios, $d_{\mathrm{p}}/\Delta x = 0.5$ and $d_{\mathrm{p}}/\Delta x = 1$, and for three different Stokes numbers, 
$\mathrm{St} = \{0.1, 1, 10 \}$. All simulations are conducted using a uniformly-spaced grid. As expected, the error in the particle trajectory and 
settling velocity is lower for the case of $d_{\mathrm{p}}/\Delta x = 0.5$. The simulations without any velocity disturbance correction predicts a 
much larger settling velocity due to the estimation of a lower drag force, resulting in a particle settling at the wall much earlier 
than in the reference simulations. This effect is more pronounced in the cases with $d_{\mathrm{p}}/\Delta x = 1$, but is still
significant for the cases with $d_{\mathrm{p}}/\Delta x = 0.5$.
The error in the normalized distance and normalized particle velocity appears to increase closest to the wall in both the cases with a wall-bounded correction, and without any correction; this is due to the use of a uniform mesh-spacing in the simulations, the 
effect of which on the grid generated velocity disturbance is more prominent close to the wall.

\section{Summary and conclusions}
\label{sec:summary}

\quad This paper presents a numerical framework to evaluate the unsteady self-induced velocity disturbance of a 
particle near a single planar wall in a fluid flow. We utilize the analytical solution from \citet{Felderhof2009}, provided in 
Fourier space for a singular impulse, and obtain the fluid velocity disturbance generated due to 
a regularized time-dependent forcing at arbitrary distances $h$ away from the wall. The velocity disturbance is obtained 
numerically by convoluting the singular Stokeslet solution with a discrete regularization kernel. 
The use of the Fourier convolution theorem and discrete Fourier transforms allows the computationally cheap 
evaluation of this regularization via a numerical convolution. The regularized Stokeslets are stored as temporal maps for various time levels. 
The maximum time level, $K_{\mathrm{max}}$, until which these temporal maps are stored is determined by the time when the 
regularized Stokeslet attains a steady state. The 
transient nature of the disturbance requires the storage of force instances that represent the momentum 
transferred from the particle to the fluid at every time instance until $K_{\mathrm{max}}$. These 
fictitous momentum sources act as one-way coupled particles and are advected with the background disturbed fluid velocity. 
The fluid velocity disturbance contributed by each force instance is the scalar product of the force with the appropriate 
vertical or horizontal regularized correction map at the corresponding time instant. The net velocity disturbance caused 
by the particle at any point in the fluid is obtained as the sum of all the velocity disturbance from all such 
fictitious momentum sources.

The presented velocity correction scheme is tested for the case of a fixed particle subjected to a uniform flow 
at different wall-normal distances, where the influence of the correction on the evaluated undisturbed velocity 
at the particle center is demonstrated and compared against corresponding simulations without any correction. A second 
test case involves a particle settling parallel to a planar wall. This scenario examines various wall-normal distances, 
Stokes numbers, particle Reynolds numbers, and particle-diameter to mesh-spacing ratios. Finally, the wall-normal motion of the 
particle is tested by subjecting a particle to an acceleration in the wall-normal direction. The settling of 
the particle, characterized by its time-dependent wall-normal distance and normalized settling velocity show 
good agreement with reference one-way coupled simulations for low to moderate Stokes numbers for 
$d_{\mathrm{p}}/\Delta x = 0.5$ and $d_{\mathrm{p}}/\Delta x = 1$ when using the newly developed model compared to 
simulations without any fluid velocity disturbance correction.

Due to the discrete convolution employed to compute the velocity disturbance, the current framework accommodates an arbitrary 
choice of regularization kernels. This capability stems from the discrete evaluation of the Fourier transform of the kernel. 
The model does not require any fitted parameters as inputs, and relies solely on physical parameters to determine 
the velocity disturbance. Since the velocity disturbance is governed by mass and momentum conservation equations, the 
resulting disturbance is a solution of the Navier-Stokes equations in the low Reynolds number limit. This approach eliminates 
the need for ad-hoc or empirical factors to describe the decay of the disturbance from the point of forcing to the wall.
The linearity of the governing equations for the velocity disturbance enables the model to address 
disturbances generated by forces in arbitrary directions. The disturbances from wall-parallel 
and wall-normal directions can be superposed to yield the resultant disturbance. Additionally, since the governing equations 
for the velocity disturbance are obtained from the volume-filtered Euler-Lagrange framework 
\cite{Evrard2024a}, the governing equations automatically account for the the local 
fluid volume fraction, making the present scheme suitable for dense particle-laden flows.

The present model can be extended to evaluate the Laplacian of the velocity disturbance, useful in computing the 
Faxen correction to the particle force in non-uniform flows \cite{Evrard2024a}. 
For particles in moderate to high Reynolds number channel flows, the disturbance generated by the particle is influenced 
by the channel walls to varying degrees. The present correction scheme can be readily implemented to evaluate the velocity 
disturbance for particles in a channel geometry when the disturbance is dictated only by a single wall.

\section{Acknowledgements}

This project has received funding from the Deutsche Forschungsgemeinschaft (DFG, German Research Foundation), 
grant number 457515061.

\section{Appendix}
\label{sec:appendix}

\quad In this section we describe the equations for the persistent Stokeslet near a planar wall. 
To obtain the persistent Stokeslet, we integrate the singular Stokeslet in time as follows,

\begin{equation}
\hat{\mathbf{V}} (k,z,h,t) = \int_0^{t} \mathbf{\hat{G}} (k,z,h,\tau) ~\mathrm{d}\tau.
\end{equation}

Here, $\mathbf{\hat{G}}$ represents the horizontal and vertical Green's functions of the unsteady Stokes 
equations near a planar wall due to a singular impulse derived in \citet{Felderhof2009}.  $\hat{\mathbf{V}}$ 
is the persistent Stokeslet in Fourier space. In the rest of this section, we 
define $\tau = \nu t$, where $\nu = \mu_{\mathrm{f}}/\rho_{\mathrm{f}}$ is the kinematic viscosity of the fluid.

\subsection{Persistent Stokeslets and their temporal asymptotic limits - Vertical forcing}
\label{subsec:appendix_persistentstokeslets_verticalforcing}

The Green's functions for a vertical forcing are divided into two parts, namely, the wall-normal velocity response $\hat{V}_{zz}$, 
and a radial velocity response, $\hat{V}_{Rz}$ respectively. $R$ is defined as the radial coordinate in the two-dimensional space 
parallel to the wall.

\subsubsection{Persistent vertical velocity response to a wall-normal forcing: ${V}_{zz}$}

We decompose the persistent vertical velocity response to a vertical forcing as follows:

\begin{align}
\begin{split}
\hat{V}_{zz} (k,z,h,t) &= 2 k^2 (\hat{V}_{zz}^{(1)} + \hat{V}_{zz}^{(2)} + \hat{V}_{zz}^{(3)} + \hat{V}_{zz}^{(4)}) + \frac{k}{2} \exp(-k(h+z)) (\hat{V}_{zz}^{(5)} + \hat{V}_{zz}^{(6)} + \hat{V}_{zz}^{(7)} + \hat{V}_{zz}^{(8)})
\\
&+ \hat{V}_{zz}^{(9)} + \frac{k}{2} \exp(k(h-z)) \hat{V}_{zz}^{(10)} + \frac{k}{2} \exp(-k(h-z)) \hat{V}_{zz}^{(12)},
\end{split}
\label{eqn:Vzz_fourier_persistent}
\end{align}

where,

\begin{align}
\begin{split}
\hat{V}_{zz}^{(1)} (k,z,h,t) &= -\exp \left(-\frac{z^2}{4\tau} - k^2 \tau - kh \right) \sqrt{\frac{t}{\pi \nu}} \frac{1}{k^2} - \frac{\exp(-k(h+z))}{4k^3\nu}
\\
& \Biggl[ (1 + kz) \erf \left( \frac{z-2k\tau}{2\sqrt{\tau}} \right) 
+ (-1 + kz) \left(1 - \exp(2kz) \erfc \left( \frac{z+2k\tau}{2\sqrt{\tau}} \right) \right) \Biggr],
\end{split}
\label{eqn:Vzz_fourier_persistent_term1}
\end{align}

\begin{equation}
\boxed{\lim_{t \to 0} \hat{V}_{zz}^{(1)} (k,z,h,t) = - 2kz \frac{\exp(-k(h+z))}{4k^3\nu},}
\label{eqn:Vzz_fourier_persistent_term1_ltt0}
\end{equation}

\begin{align}
\begin{split}
\hat{V}_{zz}^{(2)} (k,z,h,t) &= -\exp \left(-\frac{h^2}{4\tau} - k^2 \tau - kz \right) \sqrt{\frac{t}{\pi \nu}} \frac{1}{k^2}  - \frac{\exp(-k(h+z))}{4k^3\nu}
\\
& \Biggl[ (1 + kh) \erf \left( \frac{h-2k\tau}{2\sqrt{\tau}} \right) 
+ (-1 + kh) \left(1 - \exp(2kh) \erfc \left( \frac{h+2k\tau}{2\sqrt{\tau}} \right) \right) \Biggr],
\end{split}
\label{eqn:Vzz_fourier_persistent_term2}
\end{align}

\begin{equation}
\boxed{\lim_{t \to 0} \hat{V}_{zz}^{(2)} (k,z,h,t) = - 2kh \frac{\exp(-k(h+z))}{4k^3\nu},}
\label{eqn:Vzz_fourier_persistent_term2_ltt0}
\end{equation}

\begin{align}
\begin{split}
\hat{V}_{zz}^{(3)} (k,z,h,t) &= \exp \left(-\frac{(h+z)^2}{4\tau} - k^2 \tau \right) \sqrt{\frac{t}{\pi \nu}} \frac{1}{k^2} + \frac{\exp(-k(h+z))}{4k^3\nu}
\\
& \Biggl[ (1 + k(h+z)) \erf \left( \frac{h+z-2k\tau}{2\sqrt{\tau}} \right) 
\\
&+ (-1 + k(h+z)) \left(1 - \exp(2k(h+z)) \erfc \left( \frac{h+z+2k\tau}{2\sqrt{\tau}} \right) \right) \Biggr],
\end{split}
\label{eqn:Vzz_fourier_persistent_term3}
\end{align}

\begin{equation}
\boxed{\lim_{t \to 0} \hat{V}_{zz}^{(3)} (k,z,h,t) = 2k(h+z) \frac{\exp(-k(h+z))}{4k^3\nu},}
\label{eqn:Vzz_fourier_persistent_term3_ltt0}
\end{equation}

\begin{align}
\begin{split}
\hat{V}_{zz}^{(4)} (k,z,h,t) &= -\frac{2 \exp(-k(h+z))}{\nu} \Biggl[ -\sqrt{\frac{\tau}{\pi}} \frac{\exp(-k^2 \tau)}{2 k^2} + \frac{\erf(k \sqrt{\tau})}{4 k^3} \Biggr],
\end{split}
\label{eqn:Vzz_fourier_persistent_term4}
\end{align}

\begin{equation}
\boxed{\lim_{t \to 0} \hat{V}_{zz}^{(4)} (k,z,h,t) = 0,}
\label{eqn:Vzz_fourier_persistent_term4_ltt0}
\end{equation}

\begin{align}
\begin{split}
\hat{V}_{zz}^{(5)} (k,z,h,t) &= \frac{1}{\nu k^2} \Biggl[ -\frac{\erf(k\sqrt{\tau})}{2}( 2 k^2 \tau + 1)^2 + 3 \erf(k \sqrt{\tau}) 
\\
&- 5k \sqrt{\frac{\tau}{\pi}} \exp(-k^2 \tau) - 2 k^3 \exp(-k^2 \tau) \tau \sqrt{\frac{\tau}{\pi}}\Biggr],
\end{split}
\label{eqn:Vzz_fourier_persistent_term5}
\end{align}

\begin{equation}
\boxed{\lim_{t \to 0} \hat{V}_{zz}^{(5)} (k,z,h,t) = 0,}
\label{eqn:Vzz_fourier_persistent_term5_ltt0}
\end{equation}

\begin{align}
\begin{split}
\hat{V}_{zz}^{(6)} (k,z,h,t) &= \frac{1}{k} \sqrt{\frac{t}{\nu \pi}} \exp \left( -\frac{(h-2k\tau)^2}{4\tau} \right) (4 - kh + 2k^2 \tau) - t (1 - 2kh + 2k^2 \tau) \erf \left( \frac{h-2k\tau}{2\sqrt{\tau}} \right) 
\\
&+ \frac{1}{2 \nu k^2} \Biggl[ -(kh(kh -2) - 2) \erf \left( \frac{h-2k\tau}{2\sqrt{\tau}} \right) 
\\
&+ (2kh - 2) \left(1 - \exp(2kh)\erfc \left( \frac{h+2k\tau}{2\sqrt{\tau}} \right) \right) \Biggr],
\end{split}
\label{eqn:Vzz_fourier_persistent_term6}
\end{align}

\begin{equation}
\boxed{\lim_{t \to 0} \hat{V}_{zz}^{(6)} (k,z,h,t) = \frac{1}{2 \nu k^2} \left( kh(4-kh) \right),}
\label{eqn:Vzz_fourier_persistent_term6_ltt0}
\end{equation}

\begin{align}
\begin{split}
\hat{V}_{zz}^{(7)} (k,z,h,t) &= \frac{1}{k} \sqrt{\frac{t}{\nu \pi}} \exp \left( -\frac{(z-2k\tau)^2}{4\tau} \right) (4 - kz + 2k^2 \tau) - t (1 - 2kz + 2k^2 \tau) \erf \left( \frac{z-2k\tau}{2\sqrt{\tau}} \right) 
\\
&+ \frac{1}{2 \nu k^2} \Bigg[ -(kz(kz -2) - 2) \erf \left( \frac{z-2k\tau}{2\sqrt{\tau}} \right) 
\\
&+ (2kz - 2) \left(1 - \exp(2kz)\erfc \left( \frac{z+2k\tau}{2\sqrt{\tau}} \right) \right) \Biggr],
\end{split}
\label{eqn:Vzz_fourier_persistent_term7}
\end{align}

\begin{equation}
\boxed{\lim_{t \to 0} \hat{V}_{zz}^{(7)} (k,z,h,t) = \frac{1}{2 \nu k^2} \left( kz(4-kz) \right),}
\label{eqn:Vzz_fourier_persistent_term7_ltt0}
\end{equation}

\begin{align}
\begin{split}
\hat{V}_{zz}^{(8)} (k,z,h,t) &= -\frac{k}{4}  \Biggl[ \frac{4}{k^2} \sqrt{\frac{t}{\nu \pi}} \exp \left( -\frac{(h+z-2k\tau)^2}{4\tau} \right) (3 - k(h+z) + 2k^2 \tau)
\\
&+ \frac{1}{\nu k^3} \Biggl( -(-3 + 2k(h+z) (-1 + k(h+z))) \erf \left( \frac{h+z-2k\tau}{2\sqrt{\tau}} \right) 
\\
&+ (-3 + 4k(h+z)) \left(1 - \exp(2k(h+z)) \erfc \left( \frac{h+z+2k\tau}{2\sqrt{\tau}} \right) \right) \Biggr) 
\\
&+ 8 t (h+z-k\tau) \erf \left( \frac{h+z-2k\tau}{2\sqrt{\tau}} \right) \Biggr],
\end{split}
\label{eqn:Vzz_fourier_persistent_term8}
\end{align}

\begin{equation}
\boxed{\lim_{t \to 0} \hat{V}_{zz}^{(8)} (k,z,h,t) = -\frac{(h+z)}{2 \nu k} (3 - k(h+z)),}
\label{eqn:Vzz_fourier_persistent_term8_ltt0}
\end{equation}

\begin{align}
\begin{split}
\hat{V}_{zz}^{(9)} (k,z,h,t) &= \frac{\exp(-k \lvert h-z \rvert)}{4\nu k} \Biggl[ \left(1 - \exp(2k \lvert h-z \rvert) \erfc \left( \frac{\lvert h-z \rvert+2k\tau}{2\sqrt{\tau}} \right) \right) (1 - k \lvert h-z \rvert) 
\\
&- (1 + k \lvert h-z \rvert) \erf \left( \frac{\lvert h-z \rvert-2k\tau}{2\sqrt{\tau}} \right) \Biggr]
\\
&+ \frac{k \exp(k(h-z))}{2} \Biggl[ -\frac{1}{k}\sqrt{\frac{t}{\pi \nu }} \exp \left(- \frac{(h-z +2k\tau)^2}{4\tau} \right) - t \erf \left( \frac{h-z+2k\tau}{2\sqrt{\tau}} \right) \Biggr]
\\
&+ \frac{k \exp(-k(h-z))}{2} \Biggl[ -\frac{1}{k}\sqrt{\frac{t}{\pi \nu }} \exp \left(- \frac{(h-z -2k\tau)^2}{4\tau} \right) + t \erf \left( \frac{h-z-2k\tau}{2\sqrt{\tau}} \right) \Biggr],
\end{split}
\label{eqn:Vzz_fourier_persistent_term9}
\end{align}

\begin{equation}
\boxed{\lim_{t \to 0} \hat{V}_{zz}^{(9)} (k,z,h,t) = -\frac{\lvert h-z \rvert}{2 \nu} \exp(-k \lvert h-z \rvert),}
\label{eqn:Vzz_fourier_persistent_term9_ltt0}
\end{equation}

\begin{align}
\begin{split}
\hat{V}_{zz}^{(10)} (k,z,h,t) &= \frac{1}{k} \sqrt{\frac{t}{\pi \nu}}\exp \left( -\frac{(h+2k\tau)^2}{4\tau} \right) + t \erf \left( \frac{h+2k\tau}{2\sqrt{\tau}} \right) 
\\
&+ \frac{\exp(-2kh)}{4\nu k^2} \Biggl[ \erf \left( \frac{h-2k\tau}{2\sqrt{\tau}} \right) + (-1 + 2kh) \left(1 - \exp(2kh) \erfc \left( \frac{h+2k\tau}{2\sqrt{\tau}} \right) \right) \Biggr],
\end{split}
\label{eqn:Vzz_fourier_persistent_term10}
\end{align}

\begin{equation}
\boxed{\lim_{t \to 0} \hat{V}_{zz}^{(10)} (k,z,h,t) = \frac{h}{2 \nu k } \exp(-2kh),}
\label{eqn:Vzz_fourier_persistent_term10_ltt0}
\end{equation}

\begin{align}
\begin{split}
\hat{V}_{zz}^{(12)} (k,z,h,t) &= \frac{1}{k}\sqrt{\frac{t}{\pi \nu}}\exp \left( -\frac{(z+2k\tau)^2}{4\tau} \right) + t \erf \left( \frac{z+2k\tau}{2\sqrt{\tau}} \right) 
\\
&+ \frac{\exp(-2kz)}{4\nu k^2} \Biggl[ \erf \left( \frac{z-2k\tau}{2\sqrt{\tau}} \right) + (-1 + 2kz) \left(1 - \exp(2kz) \erfc \left( \frac{z+2k\tau}{2\sqrt{\tau}} \right) \right) \Biggr],
\end{split}
\label{eqn:Vzz_fourier_persistent_term12}
\end{align}

\begin{equation}
\boxed{\lim_{t \to 0} \hat{V}_{zz}^{(12)} (k,z,h,t) = \frac{z}{2 \nu k } \exp(-2kz).}
\label{eqn:Vzz_fourier_persistent_term12_ltt0}
\end{equation}


Substituting all the terms into Equation \eqref{eqn:Vzz_fourier_persistent}, we obtain the expression for the persistent Green's function $\hat{V}_{zz}$. 
Similarly, substituting the $t \to 0$ limits of each term into Equation \eqref{eqn:Vzz_fourier_persistent} gives us the following expression:

\begin{equation}
\boxed{\lim_{t \to 0} \hat{V}_{zz} (k,z,h,t) = \frac{\exp(-k(h+z))}{2 \nu} (h+z+khz) - \frac{\lvert h-z \rvert}{2 \nu} \exp(-k \lvert h-z \rvert).}
\label{eqn:Vzz_fourier_persistent_ltt0}
\end{equation}

Similarly, the $t \to \infty$ limit of $\hat{V}_{zz}$ is obtained similarly, and is given as:

\begin{equation}
\boxed{\lim_{t \to \infty} \hat{V}_{zz} (k,z,h,t) = - \frac{\exp(-k(h+z)) }{2 \nu} khz + \frac{\exp(-k \lvert h-z \rvert)}{2\nu k}.}
\label{eqn:Vzz_fourier_persistent_lttinfty}
\end{equation}

The corresponding real space Green's function is obtained via a Hankel transform using Equations \eqref{eqn:Vzz_fourier_persistent} and \eqref{eqn:Vzz_fourier_persistent_ltt0} as follows:

\begin{equation}
V_{zz} (R,z,h,t) = \frac{1}{4 \pi \rho_{\mathrm{f}}} \int_0^{\infty} \left( \hat{V}_{zz} - \lim_{t \to 0} \hat{V}_{zz} \right)J_0 (k R) k ~\mathrm{d}k.
\label{eqn:hankeltransform_Vzz}
\end{equation}

\subsubsection{Persistent horizontal velocity response to a wall-normal forcing: ${V}_{Rz}$}

The persistent radial velocity response for a vertical forcing, $\hat{V}_{Rz}$ can also be decomposed similar to Equation \eqref{eqn:Vzz_fourier_persistent}, 
and is as follows:

\begin{equation}
\hat{V}_{Rz} (k,z,h,t) = \hat{V}_{zz} + k \exp(-k(h+z)) (\hat{V}_{Rz}^{(1)} - \hat{V}_{Rz}^{(2)}) - k \exp(-k(h-z)) (\hat{V}_{Rz}^{(3)} + \hat{V}_{Rz}^{(4)}).
\label{eqn:Vrz_fourier_persistent}
\end{equation}

Here,

\begin{align}
\begin{split}
\hat{V}_{Rz}^{(1)} (k,z,h,t) &= -\frac{1}{k} \sqrt{\frac{t}{\pi \nu}}\exp \left( - \frac{(h+z-2k\tau)^2}{4\tau} \right) + \frac{1}{4 \nu k^2} \Biggl[ -(1 + 2k(h+z)) \erf \left( \frac{h+z-2k\tau}{2\sqrt{\tau}} \right) 
\\
&+ 1 - \exp(2k(h+z)) \erfc \left( \frac{h+z+2k\tau}{2\sqrt{\tau}} \right) \Biggr] + t \erf \left( \frac{h+z-2k\tau}{2\sqrt{\tau}} \right),
\end{split}
\label{eqn:Vrz_fourier_persistent_term1}
\end{align}

\begin{equation}
\boxed{\lim_{t \to 0} \hat{V}_{Rz}^{(1)} (k,z,h,t) = -\frac{(h+z)}{2 \nu k },}
\label{eqn:Vrz_fourier_persistent_term1_ltt0}
\end{equation}
\begin{align}
\begin{split}
\hat{V}_{Rz}^{(2)} (k,z,h,t) &= -\frac{1}{k} \sqrt{\frac{t}{\pi \nu}}\exp \left( - \frac{(z-2k\tau)^2}{4\tau} \right) + \frac{1}{4 \nu k^2} \Biggl[ -(1 + 2kz) \erf \left( \frac{z-2k\tau}{2\sqrt{\tau}} \right) 
\\
&+ 1 - \exp(2kz) \erfc \left( \frac{z+2k\tau}{2\sqrt{\tau}} \right) \Biggr] + t \erf \left( \frac{z-2k\tau}{2\sqrt{\tau}} \right),
\end{split}
\label{eqn:Vrz_fourier_persistent_term2}
\end{align}

\begin{equation}
\boxed{\lim_{t \to 0} \hat{V}_{Rz}^{(2)} (k,z,h,t) = -\frac{z}{2 \nu k },}
\label{eqn:Vrz_fourier_persistent_term2_ltt0}
\end{equation}


\begin{align}
\begin{split}
\hat{V}_{Rz}^{(3)} &= - \frac{1}{k} \sqrt{\frac{t}{\pi \nu}} \exp \left( -\frac{(h-z-2k\tau)^2}{4\tau} \right) + t \erf \left( \frac{h-z-2k\tau}{2\sqrt{\tau}} \right) 
\\
&+ \frac{\exp(k(h-z-\lvert h-z \rvert))}{4 \nu k^2} \Biggl[ (1 + k(h-z - \lvert h-z \rvert)) \left(1 - \exp(2k \lvert h-z \rvert) \erfc \left( \frac{\lvert h-z \rvert + 2k\tau}{2 \sqrt{\tau}} \right) \right) 
\\
&- (1 + k(h-z + \lvert h-z \rvert)) \erf \left( \frac{\lvert h-z \rvert - 2k\tau}{2 \sqrt{\tau}} \right) \Biggr],
\end{split}
\label{eqn:Vrz_fourier_persistent_term3}
\end{align}

\begin{equation}
\boxed{\lim_{t \to 0} \hat{V}_{Rz}^{(3)} = -\frac{\lvert h-z \rvert}{2 \nu k} \exp(k(h-z-\lvert h-z \rvert)),}
\label{eqn:Vrz_fourier_persistent_term3_ltt0}
\end{equation}

\begin{align}
\begin{split}
\hat{V}_{Rz}^{(4)} (k,z,h,t) &= \frac{1}{k}\sqrt{\frac{t}{\pi \nu}}\exp \left( - \frac{(z+2k\tau)^2}{4\tau} \right) + \frac{\exp(-2kz)}{4 \nu k^2} \Biggl[ \erf \left( \frac{z-2k\tau}{2\sqrt{\tau}} \right) 
\\
&+ (2kz - 1) \left( 1 - \exp(2kz) \erfc \left( \frac{z+2k\tau}{2\sqrt{\tau}} \right) \right) \Biggr] + t \erf \left( \frac{z+2k\tau}{2\sqrt{\tau}} \right),
\end{split}
\label{eqn:Vrz_fourier_persistent_term4}
\end{align}

\begin{equation}
\boxed{\lim_{t \to 0} \hat{V}_{Rz}^{(4)} (k,z,h,t) = \frac{z}{2 \nu k } \exp(-2kz).}
\label{eqn:Vrz_fourier_persistent_term4_ltt0}
\end{equation}

Substituting all the terms into Equation \eqref{eqn:Vrz_fourier_persistent}, we obtain the expression for the persistent Green's function $\hat{V}_{Rz}$. 
Similarly, substituting the $t \to 0$ limits of each term into Equation \eqref{eqn:Vrz_fourier_persistent} gives us the the following expression:

\begin{equation}
\boxed{\lim_{t \to 0} \hat{V}_{Rz} (k,z,h,t) = \frac{khz}{2 \nu} \exp(-k(h+z)).}
\label{eqn:Vrz_fourier_persistent_ltt0}
\end{equation}

Similarly, the $t \to \infty$ limit of $\hat{V}_{Rz}$ is obtained similarly, and is given as:

\begin{equation}
\boxed{\lim_{t \to \infty} \hat{V}_{Rz} (k,z,h,t) = - \frac{\exp(-k(h+z)) }{2 \nu k} (k^2 hz - kh + kz - 1) - \frac{\exp(-k(\lvert h-z \rvert))}{2 \nu k}  (k(h-z)).}
\label{eqn:Vrz_fourier_persistent_lttinfty}
\end{equation}

The corresponding real space Green's function is obtained via a Hankel transform using Equations \eqref{eqn:Vrz_fourier_persistent} and \eqref{eqn:Vrz_fourier_persistent_ltt0} as follows:

\begin{equation}
V_{Rz} (R,z,h,t) = \frac{1}{4 \pi \rho_{\mathrm{f}}} \int_0^{\infty} \left( \hat{V}_{Rz} - \lim_{t \to 0} \hat{V}_{Rz} \right)J_1 (k R) k ~\mathrm{d}k.
\label{eqn:hankeltransform_Vrz}
\end{equation}

\subsection{Persistent Stokeslets and their temporal asymptotic limits - Horizontal forcing}
\label{subsec:appendix_persistentstokeslets_horizontalforcing}

The Green's functions for a horizontal forcing are divided into three parts, namely, the $x$, $y$, and $z$ velocity responses $\hat{V}_{xx}$, $\hat{V}_{yx}$, 
and $\hat{V}_{zx}$ respectively for a forcing along the $x$ direction. The $x$ velocity response $\hat{V}_{xx}$ is itself divided into an isotropic part, $\hat{V}_{xx0}$ and an anisotropic part that depends on an angular coordinate in $\mathbb{R}^2$, and a function $\hat{V}_{xxc}$ defined in the next subsection. The isotropic component $\hat{V}_{xx0}$ does not depend on $\phi$, the orientation of the vector connecting the evaluation and forcing points in the two-dimensional plane parallel to the wall, and only depends on the two-dimensional radial distance $R$.

\subsubsection{Persistent $y$ velocity response to a wall-parallel forcing: ${V}_{yx}$}
\label{subsubsec:Vxxc_evaluation}

The function necessary to evaluate the anisotropic part of the persistent $x$ velocity response to an $x$-directional forcing, $\hat{V}_{xxc}$ can be decomposed as follows:

\begin{align}
\begin{split}
\hat{V}_{xxc} (k,z,h,t) &= \hat{V}_{zz} + k \exp(-k(h+z)) (2 \hat{V}_{xxc}^{(1)} - \hat{V}_{xxc}^{(2)} - \hat{V}_{xxc}^{(3)}) + \hat{V}_{xxc}^{(4)} 
\\
&- k \exp(-k(h-z)) \hat{V}_{xxc}^{(5)} - k \exp(k(h-z)) \hat{V}_{xxc}^{(7)},
\end{split}
\label{eqn:Vxxc_fourier_persistent}
\end{align}

where,

\begin{equation}
\hat{V}_{xxc}^{(1)} = \hat{V}_{Rz}^{(1)},
\label{eqn:Vxxc_fourier_persistent_term1}
\end{equation}

\begin{equation}
\hat{V}_{xxc}^{(2)} = \hat{V}_{Rz}^{(2)},
\label{eqn:Vxxc_fourier_persistent_term2}
\end{equation}

\begin{equation}
\hat{V}_{xxc}^{(5)} = \hat{V}_{Rz}^{(4)},
\label{eqn:Vxxc_fourier_persistent_term5}
\end{equation}

\begin{equation}
\hat{V}_{xxc}^{(7)} = \hat{V}_{zz}^{(10)},
\label{eqn:Vxxc_fourier_persistent_term7}
\end{equation}

\begin{align}
\begin{split}
\hat{V}_{xxc}^{(3)} (k,z,h,t) &= -\frac{1}{k}\sqrt{\frac{t}{\pi \nu}}\exp \left( - \frac{(h-2k\tau)^2}{4\tau} \right) + \frac{1}{4 \nu k^2} \Biggl[ -(1 + 2kh) \erf \left( \frac{h-2k\tau}{2\sqrt{\tau}} \right) 
\\
&+ 1 - \exp(2kh) \erfc \left( \frac{h+2k\tau}{2\sqrt{\tau}} \right) \Biggr] + t \erf \left( \frac{h-2k\tau}{2\sqrt{\tau}} \right),
\end{split}
\label{eqn:Vxxc_fourier_persistent_term3}
\end{align}

\begin{equation}
\boxed{\lim_{t \to 0} \hat{V}_{xxc}^{(3)} (k,z,h,t) = -\frac{h}{2 \nu k },}
\label{eqn:Vxxc_fourier_persistent_term3_ltt0}
\end{equation}

\begin{align}
\begin{split}
\hat{V}_{xxc}^{(4)} (k,z,h,t) &= 2 \sqrt{\frac{t}{\pi \nu}} \exp \left( - \frac{(h-z)^2}{4\tau} - k^2 \tau \right) - k t \exp(-k(h-z)) \erf \left( \frac{h-z-2k\tau}{2\sqrt{\tau}} \right) 
\\
&+ k t \exp(k(h-z)) \erf \left( \frac{h-z+2k\tau}{2\sqrt{\tau}} \right) 
\\
&+ \frac{\exp(-k \lvert h-z \rvert)}{2\nu k} \Biggl[ (1 + k\lvert h-z \rvert) \erf \left( \frac{\lvert h-z \rvert -2k\tau}{2\sqrt{\tau}} \right) 
\\
&+ (-1 + k\lvert h-z \rvert) \left( 1 - \exp(2k \lvert h-z \rvert) \erfc \left( \frac{\lvert h-z \rvert +2k\tau}{2\sqrt{\tau}} \right) \right) \Biggr],
\end{split}
\label{eqn:Vxxc_fourier_persistent_term4}
\end{align}

\begin{equation}
\boxed{\lim_{t \to 0} \hat{V}_{xxc}^{(4)} (k,z,h,t) = \frac{\lvert h-z \rvert}{\nu} \exp(-k \lvert h-z \rvert).}
\label{eqn:Vxxc_fourier_persistent_term4_ltt0}
\end{equation}


Substituting all the terms into Equation \eqref{eqn:Vxxc_fourier_persistent}, we obtain the expression for the persistent Green's function component $\hat{V}_{xxc}$. 
Similarly, substituting the $t \to 0$ limits of each term into Equation \eqref{eqn:Vxxc_fourier_persistent} gives us the the following expression:

\begin{equation}
\boxed{\lim_{t \to 0} \hat{V}_{xxc} (k,z,h,t) = \frac{\exp(-k(h+z))}{2 \nu} (khz-h-z) + \frac{\lvert h-z \rvert}{2 \nu} \exp(-k \lvert h-z \rvert).}
\label{eqn:Vxxc_fourier_persistent_ltt0}
\end{equation}

Similarly, the $t \to \infty$ limit of $\hat{V}_{xxc}$ is obtained similarly, and is given as:

\begin{equation}
\boxed{\lim_{t \to \infty} \hat{V}_{xxc} (k,z,h,t) = - \frac{\exp(-k(h+z)) }{2 \nu k} (k^2 hz - 2) - \frac{\exp(-k \lvert h-z \rvert)}{2\nu k}.}
\label{eqn:Vxxc_fourier_persistent_lttinfty}
\end{equation}

We can obtain the $\hat{V}_{yx}$ component of the horizontal forcing Green's function via a Hankel transform using Equations \eqref{eqn:Vxxc_fourier_persistent} and \eqref{eqn:Vxxc_fourier_persistent_ltt0} as follows:

\begin{equation}
V_{yx} (R,z,h,t) = -\frac{1}{4 \pi \rho_{\mathrm{f}}} \sin \phi \cos \phi \int_0^{\infty} \left( \hat{V}_{xxc} - \lim_{t \to 0} \hat{V}_{xxc} \right) J_2 (k R) k ~\mathrm{d}k.
\label{eqn:hankeltransform_Vyx}
\end{equation}

\subsubsection{Persistent $x$ velocity response to a wall-parallel forcing: ${V}_{xx}$}
\label{subsubsec:Vxx0_evaluation}
The isotropic part of the Green's function for an $x$-directional forcing can be decomposed as follows:

\begin{equation}
\hat{V}_{xx0} (k,z,h,t) = \frac{1}{2 \nu k } \left( \hat{V}_{xx0}^{(1)} + \hat{V}_{xx0}^{(2)} + \hat{V}_{xx0}^{(3)} + \hat{V}_{xx0}^{(4)} \right),
\label{eqn:Vxx0_fourier_persistent}
\end{equation}

where,

\begin{equation}
\hat{V}_{xx0}^{(1)} (k,z,h,t) = \exp(-k \lvert h - z \rvert) \erfc \left( \frac{\lvert h-z \rvert - 2k\tau}{2\sqrt{\tau}} \right),
\label{eqn:Vxx0_fourier_persistent_term1}
\end{equation}

\begin{equation}
\hat{V}_{xx0}^{(2)} (k,z,h,t) = -\exp(-k (h+z)) \erfc \left( \frac{h+z - 2k\tau}{2\sqrt{\tau}} \right),
\label{eqn:Vxx0_fourier_persistent_term2}
\end{equation}

\begin{equation}
\hat{V}_{xx0}^{(3)} (k,z,h,t) = -\exp(k \lvert h - z \rvert) \erfc \left( \frac{\lvert h-z \rvert + 2k\tau}{2\sqrt{\tau}} \right),
\label{eqn:Vxx0_fourier_persistent_term3}
\end{equation}

\begin{equation}
\hat{V}_{xx0}^{(4)} (k,z,h,t) = \exp(k (h+z)) \erfc \left( \frac{h+z + 2k\tau}{2\sqrt{\tau}} \right).
\label{eqn:Vxx0_fourier_persistent_term4}
\end{equation}

Substituting all the terms into Equation \eqref{eqn:Vxx0_fourier_persistent}, we obtain the isotropic part of the persistent Green's function for a horizontal $x$-directional forcing, $\hat{V}_{xx0}$. 
Similarly, substituting the $t \to 0$ limits of each term into Equation \eqref{eqn:Vxx0_fourier_persistent} gives us the the following expression:

\begin{equation}
\boxed{\lim_{t \to 0} \hat{V}_{xx0} (k,z,h,t) = 0.}
\label{eqn:Vxx0_fourier_persistent_ltt0}
\end{equation}

Similarly, the $t \to \infty$ limit of $\hat{V}_{xx0}$ is obtained similarly, and is given as:

\begin{equation}
\boxed{\lim_{t \to \infty} \hat{V}_{xx0} (k,z,h,t) = 0.}
\label{eqn:Vxx0_fourier_persistent_lttinfty}
\end{equation}

We can obtain the $\hat{V}_{xx}$ component of the horizontal forcing Green's function via a Hankel transform using Equations \eqref{eqn:Vxxc_fourier_persistent}, \eqref{eqn:Vxxc_fourier_persistent_ltt0}, and \eqref{eqn:Vxx0_fourier_persistent} as follows:

\begin{equation}
V_{xx} (R,z,h,t) = \frac{1}{4 \pi \rho_{\mathrm{f}}} \int_0^{\infty} \left[ \hat{V}_{xx0} J_0 (k R) k + \left( \hat{V}_{xxc} - \lim_{t \to 0} \hat{V}_{xxc} \right) \left( \frac{J_1 (k R)}{R} - \cos^2 \phi J_2 (k R) k \right) \right] \mathrm{d}k
\label{eqn:hankeltransform_Vxx}
\end{equation}


\subsubsection{Persistent $z$ velocity response to a wall-parallel forcing: ${V}_{zx}$}

The vertical velocity response to a horizontal $x$-directional forcing can be decomposed as follows:
\begin{equation}
\hat{V}_{zx} (k,z,h,t) = \hat{V}_{zz} + k \exp(-k(h+z)) (\hat{V}_{zx}^{(1)} - \hat{V}_{zx}^{(2)}) + k \exp(k(h-z)) (\hat{V}_{zx}^{(3)} - \hat{V}_{zx}^{(4)}),
\label{eqn:Vzx_fourier_persistent}
\end{equation}

where,

\begin{equation}
\hat{V}_{zx}^{(1)} = \hat{V}_{Rz}^{(1)},
\label{eqn:Vzx_fourier_persistent_term1}
\end{equation}

\begin{equation}
\hat{V}_{zx}^{(2)} = \hat{V}_{xxc}^{(3)},
\label{eqn:Vzx_fourier_persistent_term2}
\end{equation}

\begin{equation}
\hat{V}_{zx}^{(4)} = \hat{V}_{zz}^{(10)}.
\label{eqn:Vzx_fourier_persistent_term4}
\end{equation}

\begin{align}
\begin{split}
\hat{V}_{zx}^{(3)} (k,z,h,t) &= \frac{1}{k} \sqrt{\frac{t}{\pi \nu}}\exp \left(- \frac{(h-z+2k\tau)^2}{4\tau}  \right) + t \erf \left( \frac{h-z+2k\tau}{2\sqrt{\tau}} \right) 
\\
&+ \frac{\exp(-k(h-z + \lvert h-z \rvert))}{4\nu k^2} \Biggl[ (-1 + k(h-z + \lvert h-z \rvert)) 
\\
&\Biggl(1 - \exp(2k \lvert h-z \rvert) \erfc \left( \frac{\lvert h-z \rvert + 2k\tau}{2\sqrt{\tau}} \right) \Biggr) 
\\
&+ (1 + k(\lvert h-z \rvert - (h - z))) \erf \left( \frac{\lvert h-z \rvert - 2k\tau}{2\sqrt{\tau}} \right) \Biggr]
\end{split}
\label{eqn:Vzx_fourier_persistent_term3}
\end{align}

\begin{equation}
\boxed{\lim_{t \to 0} \hat{V}_{zx}^{(3)} (k,z,h,t) = \frac{\lvert h-z \rvert}{2\nu k} \exp(-k(h-z + \lvert h-z \rvert))}.
\label{eqn:Vzx_fourier_persistent_term3_ltt0}
\end{equation}

Substituting all the terms into Equation \eqref{eqn:Vzx_fourier_persistent}, we obtain the expression for the persistent Green's function component $\hat{V}_{zx}$. 
Similarly, substituting the $t \to 0$ limits of each term into Equation \eqref{eqn:Vzx_fourier_persistent} gives us the the following expression:

\begin{equation}
\boxed{\lim_{t \to 0} \hat{V}_{zx} (k,z,h,t) = \frac{khz}{2 \nu} \exp(-k(h+z)).}
\label{eqn:Vzx_fourier_persistent_ltt0}
\end{equation}

Similarly, the $t \to \infty$ limit of $\hat{V}_{xx0}$ is obtained similarly, and is given as:

\begin{equation}
\boxed{\lim_{t \to \infty} \hat{V}_{zx} (k,z,h,t) = - \frac{\exp(-k(h+z)) }{2 \nu k} (k^2 hz - kz + kh - 1) + \frac{\exp(-k \lvert h-z \rvert)}{2\nu k} (k(h-z)).}
\label{eqn:Vzx_fourier_persistent_lttinfty}
\end{equation}

We can obtain the $\hat{V}_{zx}$ component of the horizontal forcing Green's function via a Hankel transform using Equations \eqref{eqn:Vzx_fourier_persistent} and \eqref{eqn:Vzx_fourier_persistent_ltt0} as follows:

\begin{equation}
V_{zx} (R,z,h,t) = -\frac{1}{4 \pi \rho_{\mathrm{f}}} \cos \phi \int_0^{\infty} \left( \hat{V}_{zx} - \lim_{t \to 0} \hat{V}_{zx} \right) J_1 (k R) k ~\mathrm{d}k.
\label{eqn:hankeltransform_Vzx}
\end{equation}

\subsection{Central value of regularized transient Stokeslets}
\label{subsec:centralvalue_regstokeslets}

\quad Using the cell-centered values of the persistent Stokeslets against the cell-averaged quantity is accurate only 
upto $\mathcal{O}(\Delta x^2)$. The major contribution to the fluid velocity disturbance is from the value of the 
Stokeslet at $\mathbf{x = 0}$. In order to recover with a higher order accuracy near this region, we evaluate the 
central value of the Stokeslets derived in Sections \ref{subsec:appendix_persistentstokeslets_verticalforcing} 
and \ref{subsec:appendix_persistentstokeslets_horizontalforcing} by performing a spatial convolution of the 
Green's functions with a cylindrical Top-hat filter. The choice of cylindrical coordinates is to utilize the 
radial symmetry of the geometry in the wall-parallel direction. The radius of the cylinder is defined as $q$, 
and the height of the cylinder as $2q$. Consequently, the normalization factor for the top-hat filter, such that 
the volume integral of a uniform field in this cylindrical volume recovers the volume of a Cartesian fluid cell 
of volume $\Delta x^3$ is $2 \pi q^3$. Mathematically, the Top-hat filter in cylindrical coordinates is defined as 
follows:

\begin{equation}
\mathcal{K} (R,z) = \frac{1}{2\pi q^3}\left\{ \begin{array}{ll} 1 & R < q \wedge z \in [-q,+q], \\ 0 & R \ge q. \end{array}\right.
\label{eqn:tophat_filter_cyl_coordinates}
\end{equation}

The convolution of the transient persistent Stokeslets tensor, $\mathbf{V}$ with $\mathcal{K}(R,z)$ at $\mathbf{x} = \mathbf{r}_0$ 
in cylindrical coordinates is defined as,

\begin{equation}
\mathbf{V}_{\mathcal{K}} (\mathbf{r}_0,h,t) = \frac{1}{2\pi q^3} \int_{0}^{2 \pi} \int_{h-q}^{h+q} \int_{0}^{q} \mathbf{V}(R,z,h,t) R ~\mathrm{d}R ~\mathrm{d}z ~\mathrm{d}\phi.
\label{eqn:convolution_stokeslet_tophat}
\end{equation}

In the next subsections, we derive the central values of the persistent Stokeslets for all five components of the 
Green's functions discussed in the previous sections.

\subsubsection{Central value of the vertical velocity response to a wall-normal forcing: ${V}_{zz,\mathcal{K}}$}

Defining ${V}_{zz}$ from Equation \eqref{eqn:hankeltransform_Vzz}, and performing the convolution by substituting in 
Equation \eqref{eqn:convolution_stokeslet_tophat} we obtain:

\begin{equation}
V_{zz,\mathcal{K}} (\mathbf{r}_0,q,h,t) = \frac{1}{2\pi q^3} \int_{0}^{2 \pi} \int_{h-q}^{h+q} \int_{0}^{q} \Biggl[ \frac{1}{4 \pi \rho_{\mathrm{f}}} \int_0^{\infty} \left( \hat{V}_{zz} - \lim_{t \to 0} \hat{V}_{zz} \right)J_0 (k R) k ~\mathrm{d}k \Biggr] R ~\mathrm{d}R ~\mathrm{d}z ~\mathrm{d}\phi.
\label{eqn:convolution_stokeslet_tophat_Vzz}
\end{equation}

The above integral simplifies to,

\begin{equation}
V_{zz,\mathcal{K}} (\mathbf{r}_0,q,h,t) = \frac{1}{4 \pi \rho_{\mathrm{f}} q^2} \int_0^{\infty} \left( \hat{V}^{q}_{zz} - \lim_{t \to 0} \hat{V}^{q}_{zz} \right) J_1 (k q) ~\mathrm{d}k,
\label{eqn:convolution_stokeslet_tophat_Vzz_simplfied}
\end{equation}

where,

\begin{equation}
\hat{V}^{q}_{zz} (k,q,h,t) = \int_{h-q}^{h+q} \hat{V}_{zz} ~\mathrm{d}z.
\label{eqn:zintegral_convolution_Vzz}
\end{equation}

In the rest of this section, the notations $\exp(x)$ and $\mathrm{e}^{x}$ are used interchangeably for better readability, 
but they both denote the exponential function. $\hat{V}^{q}_{zz}$ can be decomposed similar to Equation 
\eqref{eqn:Vzz_fourier_persistent}, and the integral in Equation \eqref{eqn:zintegral_convolution_Vzz} is performed for 
each term. Therefore,


\begin{align}
\begin{split}
\hat{V}_{zz}^{q,(1)} (k,q,h,t) &= \int_{h-q}^{h+q} \hat{V}_{zz}^{(1)} (k,z,h,t) ~\mathrm{d}z
\\
&= \frac{1}{4 k^4 \nu} \Biggl[ 4 \mathrm{e}^{-kh} \mathrm{e}^{-k^2 \tau} \left( \erf \left(\frac{h-q}{2 \sqrt{\tau}}\right) - \erf \left(\frac{h+q}{2 \sqrt{\tau}}\right) \right) (k^2 \tau + 1)
\\
&+ \mathrm{e}^{-k(2h + q)} k(h+q) - \mathrm{e}^{-k(2h - q)} k(h-q) - \mathrm{e}^{-k(2h - q)} (2 + k(h-q)) \erf \left(\frac{h-q-2k\tau}{2 \sqrt{\tau}}\right)
\\
&- \mathrm{e}^{-kq} (-2 + k(h-q)) \erfc \left(\frac{h-q+2k\tau}{2 \sqrt{\tau}}\right) + \mathrm{e}^{-k(2h + q)} (2 + k(h+q)) \erf \left(\frac{h+q-2k\tau}{2 \sqrt{\tau}}\right)
\\
&+ \mathrm{e}^{kq} (-2 + k(h+q)) \erfc \left(\frac{h+q+2k\tau}{2 \sqrt{\tau}}\right) \Biggr],
\end{split}
\label{eqn:Vzz_centralvalue_persistent_term1}
\end{align}


\begin{align}
\begin{split}
\hat{V}_{zz}^{q,(2)} (k,q,h,t) &= \int_{h-q}^{h+q} \hat{V}_{zz}^{(2)} (k,z,h,t) ~\mathrm{d}z
\\
&= \frac{1}{2 k^4 \nu} \Biggl[ -2 k \sqrt{\frac{\tau}{\pi}} \exp \left( -k(h+q) - \frac{h^2}{4\tau} - k^2 \tau \right) \left(-1 + \exp(2kq) \right)
\\
&+ \mathrm{e}^{-2kh} \sinh(kq) \Biggl( -2kh + (1 + kh) \erfc \left(\frac{h - 2 k \tau}{2 \sqrt{\tau}} \right) + \mathrm{e}^{2kh} (-1 + kh) \erfc \left(\frac{h + 2 k \tau}{2 \sqrt{\tau}} \right) \Biggr) \Biggr],
\end{split}
\label{eqn:Vzz_centralvalue_persistent_term2}
\end{align}


\begin{align}
\begin{split}
\hat{V}_{zz}^{q,(3)} (k,q,h,t) &= \int_{h-q}^{h+q} \hat{V}_{zz}^{(3)} (k,z,h,t) ~\mathrm{d}z
\\
&= \frac{1}{4 k^4 \nu} \Biggl[ 4 \mathrm{e}^{-k^2 \tau} \left( \erf \left(\frac{2h+q}{2 \sqrt{\tau}}\right) - \erf \left(\frac{2h-q}{2 \sqrt{\tau}}\right) \right) (k^2 \tau + 1)
\\
&+ \mathrm{e}^{-k(2h - q)} k(2h-q) - \mathrm{e}^{-k(2h + q)} k(2h+q) 
\\
&+ \mathrm{e}^{-k(2h - q)} (2 + k(2h-q)) \erf \left(\frac{2h-q-2k\tau}{2 \sqrt{\tau}}\right)
\\
&+ \mathrm{e}^{k(2h - q)} (-2 + k(2h-q)) \erfc \left(\frac{2h-q+2k\tau}{2 \sqrt{\tau}}\right) 
\\
&- \mathrm{e}^{-k(2h + q)} (2 + k(2h+q)) \erf \left(\frac{2h+q-2k\tau}{2 \sqrt{\tau}}\right)
\\
&- \mathrm{e}^{k(2h + q)} (-2 + k(2h+q)) \erfc \left(\frac{2h+q+2k\tau}{2 \sqrt{\tau}}\right) \Biggr],
\end{split}
\label{eqn:Vzz_centralvalue_persistent_term3}
\end{align}


\begin{align}
\begin{split}
\hat{V}_{zz}^{q,(4)} (k,q,h,t) &= \int_{h-q}^{h+q} \hat{V}_{zz}^{(4)} (k,z,h,t) ~\mathrm{d}z
\\
&= -\frac{4 \mathrm{e}^{-2kh} \sinh(kq)}{k \nu} \Biggl[ -\sqrt{\frac{\tau}{\pi}} \frac{\exp(-k^2 \tau)}{2 k^2} + \frac{\erf(k \sqrt{\tau})}{4 k^3} \Biggr],
\end{split}
\label{eqn:Vzz_centralvalue_persistent_term4}
\end{align}


\begin{align}
\begin{split}
\hat{V}_{zz}^{q,(5)} (k,q,h,t) &= \int_{h-q}^{h+q} \frac{k}{2} \exp(-k(h+z)) \hat{V}_{zz}^{(5)} (k,z,h,t) ~\mathrm{d}z
\\
&= \frac{\exp(-2kh) \sinh(kq)}{\nu k^2} \hat{V}_{zz}^{(5)} (k,q,h,t),
\end{split}
\label{eqn:Vzz_centralvalue_persistent_term5}
\end{align}


\begin{align}
\begin{split}
\hat{V}_{zz}^{q,(6)} (k,q,h,t) &= \int_{h-q}^{h+q} \frac{k}{2} \exp(-k(h+z)) \hat{V}_{zz}^{(6)} (k,z,h,t) ~\mathrm{d}z
\\
&= \exp(-2kh) \sinh(kq) \hat{V}_{zz}^{(6)} (k,q,h,t),
\end{split}
\label{eqn:Vzz_centralvalue_persistent_term6}
\end{align}


\begin{align}
\begin{split}
\hat{V}_{zz}^{q,(7)} (k,q,h,t) &= \int_{h-q}^{h+q} \frac{k}{2} \exp(-k(h+z)) \hat{V}_{zz}^{(7)} (k,z,h,t) ~\mathrm{d}z
\\
&= -\frac{\exp(-kh)}{4 k^2 \nu \sqrt{\pi}} \Biggl[ 4 k \sqrt{\tau} \mathrm{e}^{-k^2 \tau} \left( \mathrm{e}^{-(h-q)^2 /4\tau} - \mathrm{e}^{-(h+q)^2 /4\tau} \right) 
\\
&+ (6 + 8k^2\tau) \mathrm{e}^{-k^2 \tau} \sqrt{\pi} \left( \erf \left( \frac{h-q}{2\sqrt{\tau}} \right) - \erf \left( \frac{h+q}{2\sqrt{\tau}} \right) \right) 
\\
&+ \mathrm{e}^{-k(h-q)} \sqrt{\pi} (-2 + k^2(-2\tau + (h-q - 2k\tau)^2)) \erf \left( \frac{h-q-2k\tau}{2\sqrt{\tau}} \right) 
\\
&- \mathrm{e}^{-k(h+q)} \sqrt{\pi} (-2 + k^2(-2\tau + (h+q - 2k\tau)^2)) \erf \left( \frac{h+q-2k\tau}{2\sqrt{\tau}} \right)
\\
&+ 2 \mathrm{e}^{k(h-q)} \sqrt{\pi} (-2 + k(h-q)) \erf \left( \frac{h-q+2k\tau}{2\sqrt{\tau}} \right)
\\
&- 2 \mathrm{e}^{k(h+q)} \sqrt{\pi} (-2 + k(h+q)) \erf \left( \frac{h+q+2k\tau}{2\sqrt{\tau}} \right)
\\
&+ 2 k^2 \sqrt{\tau} (2k\tau - (h+q)) \exp \left( - \frac{(h+q+2k\tau)^2}{4\tau} + k(h+q) \right)
\\
&- 2 k^2 \sqrt{\tau} (2k\tau - (h-q)) \exp \left( - \frac{(h-q+2k\tau)^2}{4\tau} + k(h-q) \right) 
\\
&+ 4\sqrt{\pi} (\mathrm{e}^{k(h-q)} - \mathrm{e}^{k(h+q)}) + 4 k \sqrt{\pi} ((h+q)\cosh(k(h+q)) - (h-q)\cosh(k(h-q))) \Biggr],
\end{split}
\label{eqn:Vzz_centralvalue_persistent_term7}
\end{align}

\begin{align}
\begin{split}
\hat{V}_{zz}^{q,(8)} (k,q,h,t) &= \int_{h-q}^{h+q} \frac{k}{2} \exp(-k(h+z)) \hat{V}_{zz}^{(8)} (k,z,h,t) ~\mathrm{d}z
\\
&= -\frac{1}{8 k^2 \nu} \Biggl[ 2\mathrm{e}^{-2kh} \sinh(kq) + 4\mathrm{e}^{-k(2h-q)} k(2h-q) - 4\mathrm{e}^{-k(2h+q)} k(2h+q) 
\\
&+ 4 k \sqrt{\frac{\tau}{\pi}} \mathrm{e}^{-k^2\tau} \biggl( \mathrm{e}^{-(2h+q)^2 /4\tau} (3 + k(2h+q) -2k^2\tau) 
\\
&-  \mathrm{e}^{-(2h-q)^2 /4\tau} (3 + k(2h-q) -2k^2\tau) \biggr)
\\
&+ 8 \mathrm{e}^{-k^2\tau} (1 + 2k^2\tau) \left(\erf \left( \frac{2h+q}{2\sqrt{\tau}} \right) - \erf \left( \frac{2h-q}{2\sqrt{\tau}} \right) \right)
\\
&- \mathrm{e}^{-k (2h - q)} \biggr(-1 + 2 k \biggl(4kh^2 - q + h \left(2 - 4k \left(q + 2k\tau \right) \right) 
\\
&+ k \left(-4 \tau + (q + 2 k \tau)^2 \right) \biggr) \biggr) \erf \left( \frac{2h-q-2k\tau}{2\sqrt{\tau}} \right)
\\
&+ \mathrm{e}^{-k (2h + q)} \biggl(-1 + 2 k \biggl(4kh^2 + q + h \left(2 + 4k \left(q - 2k\tau \right) \right) 
\\
&+ k \left(-4 \tau + \left(q - 2 k \tau \right)^2 \right) \biggr) \biggr) \erf \left( \frac{2h+q-2k\tau}{2\sqrt{\tau}} \right)
\\
&- (7 - 4k(2h - q)) \mathrm{e}^{k (2h - q)} \erfc \left( \frac{2h-q+2k\tau}{2\sqrt{\tau}} \right) 
\\
&+ (7 - 4k(2h + q)) \mathrm{e}^{k (2h + q)} \erfc \left( \frac{2h+q+2k\tau}{2\sqrt{\tau}} \right) \Biggr],
\end{split}
\label{eqn:Vzz_centralvalue_persistent_term8}
\end{align}

\begin{align}
\begin{split}
\hat{V}_{zz}^{q,(9)} (k,q,h,t) &= \int_{h-q}^{h+q} \hat{V}_{zz}^{(9)} (k,z,h,t) ~\mathrm{d}z
\\
&= \frac{1}{2 k^2 \nu} \Biggl[ 2 - 4 \mathrm{e}^{-k^2 \tau} \erf \left( \frac{q}{2\sqrt{\tau}} \right) 
\\
&+ 2 \mathrm{e}^{-kq} (kq + 1) \erf \left( \frac{q-2k\tau}{2\sqrt{\tau}} \right) + \mathrm{e}^{kq} (kq - 2) \erfc \left( \frac{q+2k\tau}{2\sqrt{\tau}} \right) \Biggr]
\\
&-t \Biggl[ \mathrm{e}^{-kq} \erf \left( \frac{q-2k\tau}{2\sqrt{\tau}} \right) + \mathrm{e}^{kq} \erf \left( \frac{q+2k\tau}{2\sqrt{\tau}} \right)\Biggr],
\end{split}
\label{eqn:Vzz_centralvalue_persistent_term9}
\end{align}


\begin{align}
\begin{split}
\hat{V}_{zz}^{q,(10)} (k,q,h,t) &= \int_{h-q}^{h+q} \frac{k}{2} \exp(k(h-z)) \hat{V}_{zz}^{(10)} (k,z,h,t) ~\mathrm{d}z
\\
&= \sinh(kq) \hat{V}_{zz}^{(10)} (k,z,h,t),
\end{split}
\label{eqn:Vzz_centralvalue_persistent_term10}
\end{align}

\begin{align}
\begin{split}
\hat{V}_{zz}^{q,(12)} (k,q,h,t) &= \int_{h-q}^{h+q} \frac{k}{2} \exp(-k(h-z)) \hat{V}_{zz}^{(12)} (k,z,h,t) ~\mathrm{d}z
\\
&= \frac{1}{8 k^2 \nu} \Biggl[ -4 k \sqrt{\frac{\tau}{\pi}} \mathrm{e}^{-kh} \mathrm{e}^{-k^2 \tau} \left( \mathrm{e}^{-(h-q)^2 / 4\tau} - \mathrm{e}^{-(h+q)^2 / 4\tau} \right) 
\\
&+ 4 \mathrm{e}^{-kh} \mathrm{e}^{-k^2 \tau} \left( \erf \left( \frac{h+q}{2\sqrt{\tau}} \right) - \erf \left( \frac{h-q}{2\sqrt{\tau}} \right) \right) 
\\
&+ \mathrm{e}^{-k(2h-q)} \erf \left( \frac{h-q-2k\tau}{2\sqrt{\tau}} \right) - \mathrm{e}^{-k(2h+q)} \erf \left( \frac{h+q-2k\tau}{2\sqrt{\tau}} \right) 
\\
&+ \mathrm{e}^{-kq} (3 - 4k^2 \tau - 2k(h-q)) \erf \left( \frac{h-q+2k\tau}{2\sqrt{\tau}} \right)
\\
&- \mathrm{e}^{kq} (3 - 4k^2 \tau - 2k(h+q)) \erf \left( \frac{h+q+2k\tau}{2\sqrt{\tau}} \right)
\\
&- 4 (1 + \mathrm{e}^{-2kh}) kq \cosh(kq) + 2 \sinh(kq) (\mathrm{e}^{-2kh} (1 + 2kh) - 2kh + 3) \Biggr].
\end{split}
\label{eqn:Vzz_centralvalue_persistent_term12}
\end{align}


The resulting expression for $\hat{V}^{q}_{zz}$ by using Equations \eqref{eqn:Vzz_centralvalue_persistent_term1} to \eqref{eqn:Vzz_centralvalue_persistent_term12} is as follows:

\begin{align}
\begin{split}
\hat{V}_{zz}^{q} (k,q,h,t) &= 2 k^2 (\hat{V}_{zz}^{q,(1)} + \hat{V}_{zz}^{q,(2)} + \hat{V}_{zz}^{q,(3)} + \hat{V}_{zz}^{q,(4)}) + \hat{V}_{zz}^{q,(5)} + \hat{V}_{zz}^{q,(6)} + \hat{V}_{zz}^{q,(7)} + \hat{V}_{zz}^{q,(8)}
\\
&+ \hat{V}_{zz}^{q,(9)} + \hat{V}_{zz}^{q,(10)} + \hat{V}_{zz}^{q,(12)}.
\end{split}
\label{eqn:Vzz_centralvalue_persistent}
\end{align}

We obtain the $t \to 0$ limit of the above expression as follows:

\begin{align}
\begin{split}
\lim_{t \to 0} \hat{V}_{zz}^{q} (k,q,h,t) &= \frac{1}{2 k^2 \nu} \Biggl[ -2 + 2 \mathrm{e}^{-kq} (1 + kq) - \mathrm{e}^{-k(2h + q)} (1 + k^2 h^2 + kq + kh (3 + kq) ) 
\\
&+ \mathrm{e}^{-k(2h - q)} (1 + k^2 h^2 - kq + kh (3 - kq) ) \Biggr].
\end{split}
\label{eqn:Vzz_centralvalue_persistent_ltt0}
\end{align}

Substituting Equations \eqref{eqn:Vzz_centralvalue_persistent} and \eqref{eqn:Vzz_centralvalue_persistent_ltt0} into Equation \eqref{eqn:convolution_stokeslet_tophat_Vzz_simplfied}, we obtain the central value of the real space Green's function, $V_{zz,\mathcal{K}}$.

\subsubsection{Central value of the radial velocity response to a wall-normal forcing: ${V}_{Rz,\mathcal{K}}$}

Defining $V_{Rz}$ from Equation \eqref{eqn:hankeltransform_Vrz}, and performing the convolution by substituting in 
Equation \eqref{eqn:convolution_stokeslet_tophat} we obtain:

\begin{equation}
V_{Rz,\mathcal{K}} (\mathbf{r}_0,q,h,t) = \frac{1}{2\pi q^3} \int_{0}^{2 \pi} \int_{h-q}^{h+q} \int_{0}^{q} \Biggl[ \frac{1}{4 \pi \rho_{\mathrm{f}}} \int_0^{\infty} \left( \hat{V}_{Rz} - \lim_{t \to 0} \hat{V}_{Rz} \right)J_1 (k R) k ~\mathrm{d}k \Biggr] R ~\mathrm{d}R ~\mathrm{d}z ~\mathrm{d}\phi.
\label{eqn:convolution_stokeslet_tophat_Vrz}
\end{equation}

The above integral simplifies to,

\begin{equation}
V_{Rz,\mathcal{K}} (\mathbf{r}_0,q,h,t) = \frac{1}{4 \pi \rho_{\mathrm{f}} q^3} \int_0^{\infty} \left( \hat{V}^{q}_{Rz} - \lim_{t \to 0} \hat{V}^{q}_{Rz} \right) F \left( \biggl\{ \frac{3}{2} \biggr\}, \biggl\{2,\frac{5}{2} \biggr\},-\frac{k^2 q^2}{4} \right) ~\mathrm{d}k,
\label{eqn:convolution_stokeslet_tophat_Vrz_simplfied}
\end{equation}

where $_1F_2(a,\{b_1,b_2\},c)$ is the Hypergeometric function defined as,

\begin{equation}
F \left( \biggl\lbrace \frac{3}{2} \biggr\rbrace, \biggl\{2,\frac{5}{2} \biggr\},-\frac{k^2 q^2}{4} \right) = \sum_{n=0}^{\infty} \frac{3}{2 \left( n + \frac{3}{2} \right) ~(n+1)! ~n!} \left( \frac{-k^2 q^2}{4} \right)^n,
\label{eqn:hypergeometricfunction_Vrz}
\end{equation}

and

\begin{equation}
\hat{V}^{q}_{Rz} (k,q,h,t) = \int_{h-q}^{h+q} \hat{V}_{Rz} ~\mathrm{d}z.
\label{eqn:zintegral_convolution_Vrz}
\end{equation}

$\hat{V}^{q}_{Rz}$ can be decomposed similar to Equation \eqref{eqn:Vrz_fourier_persistent}, and the integral in 
Equation \eqref{eqn:zintegral_convolution_Vrz} is performed for each term. Therefore,

\begin{align}
\begin{split}
\hat{V}_{Rz}^{q,(1)} (k,q,h,t) &= \int_{h-q}^{h+q} k \exp(-k(h+z)) \hat{V}_{Rz}^{(1)} (k,z,h,t) ~\mathrm{d}z
\\
&= - \frac{1}{4 k^2 \nu} \Biggl[ 4 k \sqrt{\frac{\tau}{\pi}} \mathrm{e}^{-k^2 \tau} \left( \mathrm{e}^{-(2h-q)^2 / 4\tau} - \mathrm{e}^{-(2h+q)^2 / 4\tau} \right) 
\\
&+ 4 \mathrm{e}^{-k^2 \tau} \left( \erf \left( \frac{2h+q}{2\sqrt{\tau}} \right) - \erf \left( \frac{2h-q}{2\sqrt{\tau}} \right) \right) 
\\
&+ \mathrm{e}^{k(2h-q)} \erf \left( \frac{2h-q+2k\tau}{2\sqrt{\tau}} \right) - \mathrm{e}^{k(2h+q)} \erf \left( \frac{2h+q+2k\tau}{2\sqrt{\tau}} \right) 
\\
&+ \mathrm{e}^{-k(2h+q)} (-3 - 4kh -2kq + 4k^2 \tau) \erf \left( \frac{2h+q-2k\tau}{2\sqrt{\tau}} \right)
\\
&- \mathrm{e}^{-k(2h-q)} (-3 - 4kh + 2kq + 4k^2 \tau) \erf \left( \frac{2h-q-2k\tau}{2\sqrt{\tau}} \right)
\\
&+ 4 \sinh(2kh) \sinh(kq) \Biggr],
\end{split}
\label{eqn:Vrz_centralvalue_persistent_term1}
\end{align}


\begin{align}
\begin{split}
\hat{V}_{Rz}^{q,(2)} (k,q,h,t) &= \int_{h-q}^{h+q} k \exp(-k(h+z)) \hat{V}_{Rz}^{(2)} (k,z,h,t) ~\mathrm{d}z
\\
&= - \frac{1}{4 k^2 \nu} \Biggl[ 4 k \sqrt{\frac{\tau}{\pi}} \mathrm{e}^{-kh} \mathrm{e}^{-k^2 \tau} \left( \mathrm{e}^{-(h-q)^2 / 4\tau} - \mathrm{e}^{-(h+q)^2 / 4\tau} \right)
\\
&+ 4 \mathrm{e}^{-kh}  \mathrm{e}^{-k^2 \tau} \left( \erf \left( \frac{h+q}{2\sqrt{\tau}} \right) - \erf \left( \frac{h-q}{2\sqrt{\tau}} \right) \right) 
\\
&- \mathrm{e}^{-k(2h-q)} (3 + 2kh - 2kq - 4k^2 \tau) \erf \left( \frac{h-q-2k\tau}{2\sqrt{\tau}} \right)
\\
&- \mathrm{e}^{-k(2h+q)} (3 + 2kh + 2kq - 4k^2 \tau) \erf \left( \frac{h+q-2k\tau}{2\sqrt{\tau}} \right)
\\
&+ \mathrm{e}^{-kq} \erf \left( \frac{h-q+2k\tau}{2\sqrt{\tau}} \right) - \mathrm{e}^{kq} \erf \left( \frac{h+q+2k\tau}{2\sqrt{\tau}} \right)
\\
&- 2 \sinh(kq) \left( \mathrm{e}^{-2kh} - 1 \right) \Biggr],
\end{split}
\label{eqn:Vrz_centralvalue_persistent_term2}
\end{align}


\begin{align}
\begin{split}
\hat{V}_{Rz}^{q,(3)} (k,q,h,t) &= \int_{h-q}^{h+q} k \exp(-k(h-z)) \hat{V}_{Rz}^{(3)} (k,z,h,t) ~\mathrm{d}z
\\
&= \frac{1}{2 k^2 \nu} \Biggl[ 2 + \mathrm{e}^{-kq} kq - 4 \mathrm{e}^{-k^2 \tau} \erf \left( \frac{q}{2\sqrt{\tau}} \right) 
\\
& - \mathrm{e}^{-kq} \left( -2 -kq + 2 k^2 \tau \right) \erf \left( \frac{q - 2k\tau}{2\sqrt{\tau}} \right)
\\
&+ \mathrm{e}^{kq} \left( -2 + kq - (-2 + kq + 2 k^2 \tau ) \erf \left( \frac{q + 2k\tau}{2\sqrt{\tau}} \right) \right) \Biggr],
\end{split}
\label{eqn:Vrz_centralvalue_persistent_term3}
\end{align}


\begin{align}
\begin{split}
\hat{V}_{Rz}^{(4)} (k,h,t) &= \int_{h-q}^{h+q} k \exp(-k(h-z)) \hat{V}_{Rz}^{(4)} (k,z,h,t) ~\mathrm{d}z
\\
&= \frac{1}{4 k^2 \nu} \Biggl[ 4 k \sqrt{\frac{\tau}{\pi}} \mathrm{e}^{-kh} \mathrm{e}^{-k^2 \tau} \left( \mathrm{e}^{-(h-q)^2/4\tau} - \mathrm{e}^{-(h+q)^2/4\tau} \right) 
\\
&- \mathrm{e}^{-kq} (-3 + 2kh - 2kq + 4k^2\tau) \erfc \left( \frac{h-q+2k\tau}{2\sqrt{\tau}} \right) 
\\
&+ \mathrm{e}^{kq} (-3 + 2kh + 2kq + 4k^2\tau) \erfc \left( \frac{h+q+2k\tau}{2\sqrt{\tau}} \right) 
\\
& + 4 \mathrm{e}^{-kh} \mathrm{e}^{-k^2 \tau} \left( \erf \left( \frac{h+q}{2\sqrt{\tau}} \right) - \erf \left( \frac{h-q}{2\sqrt{\tau}} \right) \right) + \mathrm{e}^{-k (2h - q)} \erf \left( \frac{h-q-2k\tau}{2\sqrt{\tau}} \right) 
\\
&- \mathrm{e}^{-k (2h + q)} \erf \left( \frac{h+q-2k\tau}{2\sqrt{\tau}} \right) \Biggr] 
\\
&+ 2t \sinh(kq) + \frac{\mathrm{e}^{-2kh} (\sinh(kq) (1+2kh) - 2 kq \cosh(kq)) }{2 k^2 \nu}.
\end{split}
\label{eqn:Vrz_centralvalue_persistent_term4}
\end{align}

The resulting expression for $\hat{V}^{q}_{Rz}$ by using Equations \eqref{eqn:Vrz_centralvalue_persistent_term1} to \eqref{eqn:Vrz_centralvalue_persistent_term4} is as follows:

\begin{equation}
\hat{V}_{Rz}^{q} (k,q,h,t) = \hat{V}_{zz} + \hat{V}_{Rz}^{q,(1)} - \hat{V}_{Rz}^{q,(2)} - \hat{V}_{Rz}^{q,(3)} - \hat{V}_{Rz}^{q,(4)}.
\label{eqn:Vrz_centralvalue_persistent}
\end{equation}

We obtain the $t \to 0$ limit of the above expression as follows:

\begin{align}
\begin{split}
\lim_{t \to 0} \hat{V}_{Rz}^{q} (k,q,h,t) &= \frac{h \exp(-2kh)}{\nu k} \Biggl[ (1 + kh) \sinh(kq) - kq \cosh(kq) \Biggr].
\end{split}
\label{eqn:Vrz_centralvalue_persistent_ltt0}
\end{align}

Substituting Equations \eqref{eqn:Vrz_centralvalue_persistent} and \eqref{eqn:Vrz_centralvalue_persistent_ltt0} into Equation \eqref{eqn:convolution_stokeslet_tophat_Vrz_simplfied}, we obtain the central value of the real space Green's function, $V_{Rz,\mathcal{K}}$.

\subsubsection{Central value of the $y$ velocity response to a wall-parallel forcing: ${V}_{yx,\mathcal{K}}$}

Defining $V_{yx}$ from Equation \eqref{eqn:hankeltransform_Vyx}, and performing the convolution by substituting in 
Equation \eqref{eqn:convolution_stokeslet_tophat} we obtain:

\begin{equation}
V_{yx,\mathcal{K}} (\mathbf{r}_0,q,h,t) = - \frac{1}{2\pi q^3} \int_{0}^{2 \pi} \sin \phi \cos \phi \int_{h-q}^{h+q} \int_{0}^{q} \Biggl[ \frac{1}{4 \pi \rho_{\mathrm{f}}} \int_0^{\infty} \left( \hat{V}_{xxc} - \lim_{t \to 0} \hat{V}_{xxc} \right)J_2 (k R) k ~\mathrm{d}k \Biggr] R ~\mathrm{d}R ~\mathrm{d}z ~\mathrm{d}\phi.
\label{eqn:convolution_stokeslet_tophat_Vyx}
\end{equation}

Since $\int_{0}^{2 \pi} \sin \phi \cos \phi = 0$, the central value of the the $y$ velocity response is identically zero. Therefore,

\begin{equation}
V_{yx,\mathcal{K}} (\mathbf{r}_0,q,h,t) = 0.
\label{eqn:convolution_stokeslet_tophat_Vyx_simplfied}
\end{equation}

\subsubsection{Central value of the $z$ velocity response to a wall-parallel forcing: ${V}_{zx,\mathcal{K}}$}

Defining $V_{zx}$ from Equation \eqref{eqn:hankeltransform_Vzx}, and performing the convolution by substituting in 
Equation \eqref{eqn:convolution_stokeslet_tophat} we obtain:

\begin{equation}
V_{zx,\mathcal{K}} (\mathbf{r}_0,q,h,t) = - \frac{1}{2\pi q^3} \int_{0}^{2 \pi} \cos \phi \int_{h-q}^{h+q} \int_{0}^{q} \Biggl[ \frac{1}{4 \pi \rho_{\mathrm{f}}} \int_0^{\infty} \left( \hat{V}_{zx} - \lim_{t \to 0} \hat{V}_{zx} \right)J_1 (k R) k ~\mathrm{d}k \Biggr] R ~\mathrm{d}R ~\mathrm{d}z ~\mathrm{d}\phi.
\label{eqn:convolution_stokeslet_tophat_Vzx}
\end{equation}

Since $\int_{0}^{2 \pi} \cos \phi = 0$, the central value of the the $z$ velocity response is identically zero. Therefore,

\begin{equation}
V_{zx,\mathcal{K}} (\mathbf{r}_0,q,h,t) = 0.
\label{eqn:convolution_stokeslet_tophat_Vzx_simplfied}
\end{equation}

\subsubsection{Central value of the $x$ velocity response to a wall-parallel forcing: ${V}_{xx,\mathcal{K}}$}

Defining $V_{xx}$ from Equation \eqref{eqn:hankeltransform_Vxx}, and performing the convolution by substituting in 
Equation \eqref{eqn:convolution_stokeslet_tophat} we obtain:

\begin{align}
\begin{split}
V_{xx,\mathcal{K}} (\mathbf{r}_0,q,h,t) &= \frac{1}{2\pi q^3} \int_{0}^{2 \pi} \int_{h-q}^{h+q} \int_{0}^{q} \frac{1}{4 \pi \rho_{\mathrm{f}}} \int_0^{\infty} \Biggl[ \hat{V}_{xx0} J_0 (k R) k 
\\
&+ \left( \hat{V}_{xxc} - \lim_{t \to 0} \hat{V}_{xxc} \right) \left( \frac{J_1 (k R)}{R} - \cos^2 \phi J_2 (k R) k \right) \Biggr] R ~\mathrm{d}k ~\mathrm{d}R ~\mathrm{d}z ~\mathrm{d}\phi.
\end{split}
\label{eqn:convolution_stokeslet_tophat_Vxx}
\end{align}

The above integral simplifies to,

\begin{align}
\begin{split}
V_{xx,\mathcal{K}} (\mathbf{r}_0,q,h,t) &= \frac{1}{8 \pi \rho_{\mathrm{f}} q^3} \int_0^{\infty} \Biggl[2 \hat{V}_{xx0}^{q} + \left( \hat{V}_{xxc}^{q} - \lim_{t \to 0} \hat{V}_{xxc}^{q} \right) \Biggr] J_1 (k q) q ~\mathrm{d}k ~\mathrm{d}z,
\end{split}
\label{eqn:convolution_stokeslet_tophat_Vxx_simplified}
\end{align}

where,

\begin{equation}
\hat{V}^{q}_{xx0} (k,q,h,t) = \int_{h-q}^{h+q} \hat{V}_{xx0} ~\mathrm{d}z, \quad \textnormal{and} \quad \hat{V}^{q}_{xxc} (k,q,h,t) = \int_{h-q}^{h+q} \hat{V}_{xxc} ~\mathrm{d}z.
\label{eqn:zintegral_convolution_Vxz}
\end{equation}

Similar to Section \ref{subsubsec:Vxxc_evaluation}, the central value of the anisotropic component, $\hat{V}^{q}_{xxc}$ can be decomposed as follows:

\begin{align}
\begin{split}
\hat{V}^{q}_{xxc} (k,z,h,t) &= \hat{V}^{q}_{zz} + 2 \hat{V}_{xxc}^{q,(1)} - \hat{V}_{xxc}^{q,(2)} - \hat{V}_{xxc}^{q,(3)} + \hat{V}_{xxc}^{q,(4)} -  \hat{V}_{xxc}^{q,(5)} - \hat{V}_{xxc}^{q,(7)}.
\end{split}
\label{eqn:Vxxc_centralvalue_persistent}
\end{align}

In the above expression, each of the terms are as follows:

\begin{equation}
\hat{V}_{xxc}^{q,(1)} = \hat{V}_{Rz}^{q,(1)},
\label{eqn:Vxxc_centralvalue_persistent_term1}
\end{equation}

\begin{equation}
\hat{V}_{xxc}^{q,(2)} = \hat{V}_{Rz}^{q,(2)},
\label{eqn:Vxxc_centralvalue_persistent_term5}
\end{equation}

\begin{equation}
\hat{V}_{xxc}^{q,(5)} = \hat{V}_{Rz}^{q,(4)},
\label{eqn:Vxxc_centralvalue_persistent_term2}
\end{equation}

\begin{equation}
\hat{V}_{xxc}^{q,(7)} = 2 \hat{V}_{zz}^{q,(10)},
\label{eqn:Vxxc_centralvalue_persistent_term7}
\end{equation}

\begin{align}
\begin{split}
\hat{V}_{xxc}^{q,(3)} (k,q,h,t) &= \int_{h-q}^{h+q} k \exp(-k(h+z)) \hat{V}_{xxc}^{(3)} ~\mathrm{d}z
\\
&= -\frac{1}{4k^2 \nu} \mathrm{e}^{-k(2h + q)} (-1 + \mathrm{e}^{2kq}) \Biggl[-1 + 4 k \sqrt{\frac{\tau}{\pi}} \mathrm{e}^{-k^2\tau} \mathrm{e}^{- h^2/4\tau - kh}
\\
&+ (1 + 2kh - 4k^2 \tau) \erf \left( \frac{h-2k\tau}{2\sqrt{\tau}} \right) + \mathrm{e}^{2kh} \erfc \left( \frac{h+2k\tau}{2\sqrt{\tau}} \right) \Biggr],
\end{split}
\label{eqn:Vxxc_centralvalue_persistent_term3}
\end{align}

\begin{align}
\begin{split}
\hat{V}_{xxc}^{q,(4)} (k,h,t) &= \int_{h-q}^{h+q} \hat{V}_{xxc}^{(4)} (k,z,h,t) ~\mathrm{d}z
\\
&= \frac{1}{k^2 \nu} \Biggl[ -\mathrm{e}^{-kq} (2 \mathrm{e}^{kq} + kq) + 4 \mathrm{e}^{-k^2 \tau} \erf \left( \frac{q}{2\sqrt{\tau}} \right) 
\\
&- \mathrm{e}^{-kq} (2 + kq) \erf \left( \frac{q - 2k\tau}{2\sqrt{\tau}} \right) - \mathrm{e}^{kq} (-2 + kq) \erfc \left( \frac{q + 2k\tau}{2\sqrt{\tau}} \right)\Biggr] 
\\
&+ 2t \Biggl[ \mathrm{e}^{kq} \erf \left( \frac{q + 2k\tau}{2\sqrt{\tau}} \right) + \mathrm{e}^{-kq} \erf \left( \frac{q - 2k\tau}{2\sqrt{\tau}} \right) \Biggr].
\end{split}
\label{eqn:Vxxc_centralvalue_persistent_term4}
\end{align}

Substituting Equations \eqref{eqn:Vxxc_centralvalue_persistent_term1} to \eqref{eqn:Vxxc_centralvalue_persistent_term4} into Equation \eqref{eqn:Vxxc_centralvalue_persistent} gives us the central value of the component $\hat{V}^{q}_{xxc}$. Similarly, the isotropic 
component, $\hat{V}_{xx0}^{q}$ can be decomposed as follows:

\begin{equation}
\hat{V}^{q}_{xx0} (k,q,h,t) = \frac{1}{2 \nu k^2 } \left( \hat{V}_{xx0}^{q,(1)} + \hat{V}_{xx0}^{q,(2)} + \hat{V}_{xx0}^{q,(3)} + \hat{V}_{xx0}^{q,(4)} \right),
\label{eqn:Vxx0_centralvalue_persistent}
\end{equation}

where,

\begin{align}
\begin{split}
\hat{V}_{xx0}^{q,(1)} (k,q,h,t) &= 4 + 2 \mathrm{e}^{-k^2 \tau} \Biggl[ \erf \left( \frac{2h + q}{2\sqrt{\tau}} \right) - \erf \left( \frac{2h - q}{2\sqrt{\tau}} \right) \Biggr],
\end{split}
\label{eqn:Vxx0_centralvalue_persistent_term1}
\end{align}

\begin{align}
\begin{split}
\hat{V}_{xx0}^{q,(2)} (k,q,h,t) &= -2 \mathrm{e}^{-kq} \erfc \left( \frac{q - 2k\tau}{2\sqrt{\tau}} \right) - 2 \mathrm{e}^{kq} \erfc \left( \frac{q + 2k\tau}{2\sqrt{\tau}} \right),
\end{split}
\label{eqn:Vxx0_centralvalue_persistent_term2}
\end{align}

\begin{align}
\begin{split}
\hat{V}_{xx0}^{q,(3)} (k,q,h,t) &= - \mathrm{e}^{k (2h-q)} \erfc \left( \frac{2h - q + 2k\tau}{2\sqrt{\tau}} \right) - \mathrm{e}^{-k(2h-q)} \erfc \left( \frac{2h - q - 2k\tau}{2\sqrt{\tau}} \right),
\end{split}
\label{eqn:Vxx0_centralvalue_persistent_term3}
\end{align}

\begin{align}
\begin{split}
\hat{V}_{xx0}^{q,(4)} (k,q,h,t) &= \mathrm{e}^{k (2h+q)} \erfc \left( \frac{2h + q + 2k\tau}{2\sqrt{\tau}} \right) + \mathrm{e}^{-k(2h+q)} \erfc \left( \frac{2h + q - 2k\tau}{2\sqrt{\tau}} \right).
\end{split}
\label{eqn:Vxx0_centralvalue_persistent_term4}
\end{align}

Substituting Equations \eqref{eqn:Vxx0_centralvalue_persistent_term1} to \eqref{eqn:Vxx0_centralvalue_persistent_term4} into Equation \eqref{eqn:Vxx0_centralvalue_persistent} gives us the central value of the component $\hat{V}^{q}_{xx0}$. By substituting Equations 
\eqref{eqn:Vxxc_centralvalue_persistent} and \eqref{eqn:Vxx0_centralvalue_persistent} into Equation \eqref{eqn:convolution_stokeslet_tophat_Vxx_simplified} 
gives us the central value $V_{xx,\mathcal{K}}$.

\bibliographystyle{elsarticle-num-names}

\end{document}